\documentclass[twocolumn,english,superscriptaddress,prb,preprintnumbers,showpacs,floatfix]{revtex4-1}
\usepackage[T1]{fontenc}
\usepackage[latin9]{inputenc}
\usepackage{color}
\usepackage{verbatim}
\usepackage{amsmath}
\usepackage{graphicx}
\usepackage{amssymb}
\usepackage{esint}

\makeatletter
\@ifundefined{textcolor}{}
 {%
   \definecolor{BLACK}{gray}{0}
   \definecolor{WHITE}{gray}{1}
   \definecolor{RED}{rgb}{1,0,0}
   \definecolor{GREEN}{rgb}{0,1,0}
   \definecolor{BLUE}{rgb}{0,0,1}
   \definecolor{CYAN}{cmyk}{1,0,0,0}
   \definecolor{MAGENTA}{cmyk}{0,1,0,0}
   \definecolor{YELLOW}{cmyk}{0,0,1,0}
 }

\usepackage{ifthen}

\def\ket#1{\vert#1\rangle}

\begin{document}

\title{Linear-scaling DFT+U with full local orbital optimization}

\author{David D. O'Regan}
\email{david.oregan@epfl.ch}
\affiliation{Cavendish Laboratory, University of Cambridge, J. J. Thomson Avenue,
Cambridge CB3 0HE, United Kingdom}
\affiliation{	Theory and Simulation of Materials, 
\'{E}cole Polytechnique F\'{e}d\'{e}rale de Lausanne, 
1015 Lausanne, Switzerland}

\author{Nicholas D. M. Hine}
\affiliation{Cavendish Laboratory, University of Cambridge, J. J. Thomson Avenue,
Cambridge CB3 0HE, United Kingdom}
\affiliation{The Thomas Young Centre and
the Department of Materials, Imperial College London, London SW7 2AZ,
United Kingdom}

\author{ Mike C. Payne}
\affiliation{Cavendish Laboratory, University of Cambridge, J. J. Thomson Avenue,
Cambridge CB3 0HE, United Kingdom}

\author{Arash A. Mostofi}
\affiliation{The Thomas Young Centre and
the Department of Materials, Imperial College London, London SW7 2AZ,
United Kingdom}

\date{\today}
\begin{abstract}
We present an approach to the 
DFT+$U$ method (Density Functional
Theory + Hubbard model) within which the 
computational effort for calculation of ground state 
energies and forces scales linearly with system size.
We employ a formulation of the Hubbard model 
using nonorthogonal projector functions 
to define the localized subspaces, 
and apply it to a local-orbital DFT method 
including \emph{in situ} orbital optimization. 
The resulting approach thus combines 
linear-scaling and systematic variational convergence. We
demonstrate the scaling of the method by applying it to nickel oxide
nano-clusters with sizes exceeding $7,000$ atoms.
\end{abstract}


\maketitle


\section{Introduction}


The success of the
Kohn-Sham formulation of 
density functional theory 
(DFT)~\cite{PhysRev.136.B864,
PhysRev.140.A1133} 
is largely owed to its capability of accurately 
and reliably reproducing the ground-state
properties of quantum-mechanical systems. 
Two factors that limit the applicability of DFT are 
the computational expense of 
treating large systems, and the difficulties
encountered in simulating so-called ``strongly-correlated''
systems. 
The realistic study of many technologically and biologically
important structures requires the explicit treatment of very large
system sizes, yet the asymptotic scaling of 
conventional DFT algorithms 
is cubic in the number of atoms, so that the 
feasible size limit for routine 
calculations is typically around $1,000$ atoms even on
powerful high-performance computing architectures. 
Only by using DFT codes for which the effort increases linearly
with system size, recently reviewed in Ref.~\onlinecite{bowlerreview}
and examples of which include 
ONETEP~\cite{10.1063/1.1839852,Hine20091041,hine:114111}, 
OPENMX~\cite{PhysRevB.72.045121,PhysRevB.82.075131}, and 
CONQUEST~\cite{PSSB:PSSB200541386,0953-8984-22-7-074207}, 
we may routinely bring first-principles simulation to bear on
pertinent technological, environmental and medical problems.
Furthermore, for many functional materials, 
most typically comprising open-shell first-row
transition metal or lanthanoid ions,
DFT with local or semi-local functionals
performs very poorly, failing to obtain qualitative 
agreement with experimental observations in the most
severe cases.
Many methods have been developed to overcome this deficiency, and
here we focus on the 
DFT+$U$~\cite{PhysRevB.44.943,*PhysRevB.48.16929}
technique due to its widespread adoption and its
amenability to linear-scaling implementation.

In this article, we present a computational methodology to tackle the
obstacles of large system size and strong correlation effects simultaneously.
Working in the framework of a linear-scaling implementation of DFT
in order to tackle the issue of system size, we fully detail a DFT+$U$
implementation, including self-consistent total-energies and forces,
and demonstrate computational scaling tests on a strongly interacting
oxide system of over $7,000$ atoms. 
Previous linear-scaling or otherwise 
large-scale implementations of DFT+$U$, 
examples including Refs.~\onlinecite{PhysRevB.70.235209,
PhysRevB.73.045110,PhysRevB.76.155128}, 
have relied on a basis of fixed user-defined 
or numerically pre-solved atomic orbitals. 
A noteworthy advantage of our approach is that 
we allow for the optimization of the orbitals representing
the Kohn-Sham density-matrix \emph{in situ}, that is during the process
of total-energy minimization, with respect to an underlying,
systematic plane-wave basis~\cite{PhysRevB.66.035119,10.1063/1.1839852}. 
In this manner, we move beyond the fixed-orbital approximation
to linear-scaling DFT and DFT+$U$. 
Using this approach, truly first-principles simulations may be 
carried out on systems comprising both strong electronic interactions
and large spatial disorder, examples including layered transition-metal
and lanthanoid oxide structures, catalytic surfaces, 
molecular magnets and organometallic biomolecules.

The article is organized as follows.
We describe the DFT+$U$ technique and its
generalization to the nonorthogonal case in Section~\ref{dftuintro}, 
after which we introduce linear-scaling DFT and define 
the notation and sparse matrix  algebra for 
linear-scaling DFT+$U$ in Section~\ref{ondftu}.
Minimization of the DFT+$U$ total-energy with 
respect to the density-matrix is detailed in Section~\ref{dm}.
The method is applied to nickel oxide clusters 
exceeding $7,000$ atoms in 
Section~\ref{nickeloxidetests}, and linear-scaling
performance is demonstrated. Following 
some concluding remarks, we detail the method used
to preserve the density-matrix idempotency and normalization
in Appendices~\ref{densitykernelvariations} and 
\ref{ngwfvariations}, 
and to compute the DFT+$U$ ionic 
forces in Appendix~\ref{ionicvariations}.

\section{DFT+$U$ method for nonorthogonal projectors}
\label{dftuintro}

The use of approximations such as the local spin-density approximation 
(LSDA)~\cite{PhysRevB.23.5048}, 
for the exchange-correlation (XC)
functional in Kohn-Sham DFT,
is appropriate and highly successful in systems where the magnitude
of each electron's kinetic energy $t$ is 
large compared with the Coulomb interaction $U$ acting on it.
In such systems, usually comprising 
elements whose $3d$ or $4f$ atomic-like
states are either completely empty or filled,
the LSDA typically provides a 
good qualitative description of both the ground-state density
and the insulating gap.
In strongly correlated systems
such as Mott-Hubbard insulators~\cite{mott}, 
however, these states are localized, partially occupied, 
and do not fall in the regime
of $U\ll t$. In such cases, the LSDA may thus perform very poorly
unless it is corrected.
The DFT+$U$ method~\cite{PhysRevB.44.943,*PhysRevB.48.16929} 
reintroduces the explicit Coulomb interaction terms,
and thus the appropriate derivative discontinuity
with respect to electronic occupation number, 
 to the approximate XC functional.  

In the DFT + Hubbard $U$ method 
(DFT+$U$)~\cite{PhysRevB.44.943,*PhysRevB.48.16929}, 
a number of spatially localized subspaces, sites labeled $I$,
wherein the $U\ll t$ regime is not expected to hold,
are selected for supplementation
with explicit Coulomb correlations beyond the LSDA level, 
retaining the bare, inexpensive 
XC functional for the remainder of the system.
The strongly interacting subspaces are 
spanned by sets of localized orbitals, 
termed the Hubbard projectors $\{  \varphi_m^{(I)}  \}$.
The selection of Hubbard projectors is a topic
of interest in itself and possible choices include
localized Wannier functions built
from the Kohn-Sham eigenfunctions 
according to maximal localization~\cite{PhysRevLett.89.167204,
PhysRevB.75.224408}, energy 
downfolding~\cite{PhysRevLett.96.166401} or
maximal Coulomb repulsion~\cite{PhysRevB.77.085122} criteria, or
indeed a total-energy minimization criterion 
in combination with a self-consistency scheme, as we have
proposed in a related article, 
Ref.~\onlinecite{PhysRevB.82.081102}.
In the description of our linear-scaling method, here we
assume only that the projectors are confined to a spatial
region, real-valued, and expressed in the same underlying, systematic basis 
as the orbitals representing the Kohn-Sham density-matrix (in
the present case a truncated set of plane waves). 

In the tensorial representation, developed in order to maintain the 
tensorial invariance of subspace occupancies, 
moments, ionic forces and the total-energy~\cite{PhysRevB.83.245124}, 
localized Hubbard projector duals are
defined by\footnote{We employ the Einstein
convention, where pairs of identical 
indices are summed over unless 
in parentheses.}
\begin{equation}
O_{mm'}^{(I)}=\langle\varphi_{m}^{(I)}\vert\varphi_{m'}^{(I)}\rangle;
\quad
\lvert \varphi^{(I)m}\rangle= \lvert 
\varphi_{m'}^{(I)}\rangle O^{(I)m'm}, 
\label{localmetric}
\end{equation}
where, by definition,
$ O^{\left( I \right)}_{mm''}
O^{\left( I \right)m''m'}=\delta_{m}^{\;\; m'}$,
such that an individual metric tensor $O^{(I)}$  
is generated and used for each subspace.
The occupancy matrix is then 
most conveniently expressed
as a mixed tensor (specifically a tensor with one contravariant 
index and one covariant index), following 
Refs.~\onlinecite{PhysRevB.83.245124,Artacho},
so that its trace is a tensorial invariant, as per
\begin{align}
\label{natrep}
n_{\quad\quad\;\; m'}^{(I)(\sigma)m}=
\langle\varphi^{(I)m}|\hat{\rho}^{(\sigma)}|
{\varphi_{m'}^{(I)}}\rangle.
\end{align}
Here, $\hat{\rho}^{(\sigma)}$ is the single-particle density-matrix
for electrons of spin $\sigma$,
formally defined by
\begin{align}
\label{Eq:DM}
\hat{\rho}^{(\sigma)}= \sum_i \lvert \psi_{i}^{(\sigma)}\rangle 
f_{i}^{(\sigma)}\langle\psi_{i}^{(\sigma)} \rvert, \end{align}
where $f_{i}^{(\sigma)}$ is the occupancy of the
Kohn-Sham orbital $\lvert \psi_{i}^{(\sigma)}\rangle$.
Using this definition, we can cast the rotationally-invariant DFT+$U$
functional of Refs.~\onlinecite{0953-8984-9-4-002, PhysRevB.52.R5467} 
into a more general, tensorially invariant form -- that is invariant
under arbitrary linear combinations of the Hubbard projectors for
a given site, following Ref.~\onlinecite{PhysRevB.83.245124}. 
Specifically, we use a
tensorially invariant generalization of the 
widely-used, simplified DFT+$U$ functional of
Ref.~\onlinecite{PhysRevB.71.035105}, where 
the energy functional is given by $E_{DFT+U} =
 E_{DFT}+E_{U}$, with
\begin{align}
E_{U}=\sum_{I\sigma}\frac{U^{(I)}}{2}\left(n_{\quad\quad\;\;
    m}^{(I)(\sigma)m}-n_{\quad\quad\;\;
    m}^{(I)(\sigma)m'}n_{\quad\quad\;\;
    m'}^{(I)(\sigma)m}\right),
\label{Eq:DFTU}
\end{align}
and $U^{(I)}$ is the screened subspace-averaged 
Coulomb repulsion.
The DFT+$U$ penalty functional approximately 
emulates the exact exchange-correlation functional by introducing a
derivative discontinuity in the total-energy with respect to the occupancy
matrix, in effect approximately enforcing the unrestricted Hartree-Fock approximation
within the subspaces $I$.

\section{linear-scaling DFT+$U$}
\label{ondftu}

We now describe the steps necessary to perform
DFT+$U$ calculations with linear-scaling expense.
We have previously demonstrated the features of
projector self-consistency~\cite{PhysRevB.82.081102} and
tensorial invariance~\cite{PhysRevB.83.245124},
in our implementation of the method in the 
ONETEP code~\cite{10.1063/1.1839852,Hine20091041}, 
however the method described in this article is applicable to
linear-scaling DFT methods~\cite{bowlerreview} generally.
The method is rigorously general to the case of nonorthogonality 
of both the local orbitals and the
Hubbard projectors.

\subsection{Framework and notation}
 
Linear-scaling DFT revolves around the optimization not of the eigenstates
of the Kohn-Sham Hamiltonian, but of the density-matrix
of Eqn.~\ref{Eq:DM} expressed in terms of a set of nonorthogonal local 
orbitals, $\{|\phi_{\alpha}\rangle\}$ (known in the ONETEP code
as Nonorthogonal Generalized Wannier Functions, 
NGWFs~\cite{PhysRevB.66.035119}), that is 
\begin{align}
\label{localdenman}
\hat{\rho}^{(\sigma)}=
\lvert \phi_{\alpha}\rangle K^{(\sigma)\alpha\beta}\langle\phi_{\beta} \rvert, 
\end{align}
where the tensor $K$ is known as the \emph{density kernel}
and is generally non-diagonal.
The exponential spatial localization of the density matrix for 
insulating materials~\cite{PhysRevLett.98.046402},  
\begin{equation}
\rho^{(\sigma)} \left( \mathbf{r}, \mathbf{r'} \right) 
= \langle \mathbf{r} \rvert \hat{\rho}^{(\sigma)} \lvert  \mathbf{r'} \rangle
\sim
\exp \left( - \gamma \lvert \mathbf{r} - \mathbf{r'} 
\rvert \right),
\end{equation}
must be exploited to achieve linear-scaling, by strictly
limiting the spatial extent of the local orbitals 
and truncating the density kernel 
to an appropriate length-scale.

The contravariant duals of the local 
orbitals are denoted by $\{|\phi^{\alpha}\rangle\}$,
and the contravariant metric on the orbitals is  the inverse
of the covariant metric $S$, so that \begin{equation}
S_{\alpha\beta}=\langle\phi_{\alpha}\vert\phi_{\beta}\rangle,\quad
S^{\alpha\beta}=\langle\phi^{\alpha}\vert\phi^{\beta}\rangle=
\left(S^{-1}\right)^{\alpha\beta}.\end{equation}
We emphasize that the metric $S$ on the local orbitals, generally a
large matrix containing information about the entire system, 
and the individual metric
$O^{\left(I\right)}$ on each DFT+$U$ subspace,
a small matrix (usually $5\times5$ for $3d$-type subspaces) and
a localized quantity, are distinct even 
in cases where the Hubbard projectors are selected 
as a proper subset of the local orbitals.

The occupation matrix for each subspace 
is expressed in terms of local orbital matrix elements
by inserting the expansion of the
density-matrix, Eq.~\ref{localdenman}, into the natural
occupancy representation of Eq.~\ref{natrep}.
Making use of the transformation rules for Hubbard
projectors given by Eq.~\ref{localmetric}, we find that
\begin{align}
n_{\quad\quad\;\; m'}^{(I)(\sigma)m}  
\label{Eq:occupmat}
=  O^{(I)mm''}\langle\varphi_{m''}^{(I)}|{\phi}_{\alpha}\rangle 
K^{(\sigma)\alpha\beta}\langle{\phi}_{\beta}|\varphi_{m'}^{(I)}\rangle.
\end{align}
We denote the overlap between Hubbard projectors 
$\{|\varphi_{m}^{(I)}\rangle\}$ and 
local orbitals $\{|\phi_{\alpha}\rangle\}$ by
\begin{equation}
V_{\beta m}^{(I)}=\langle\phi_{\beta}\vert\varphi_{m}^{(I)}\rangle;
\quad
W_{m\alpha}^{(I)}=V_{\alpha m}^{(I)\dagger}=
\langle\varphi_{m}^{(I)}|\phi_{\alpha}\rangle,
\end{equation}
 which may be very sparse matrices, particularly for a
low density of subspaces.
The DFT+$U$ correction to the total-energy, given by
Eqn.~\ref{Eq:DFTU}, is then computed with linear-scaling
cost using the sparse matrix trace
\begin{align}
\label{compactenergy}
E_{U}={} & \sum_{I,\sigma}\frac{U^{(I)}}{2}{\rm Tr} 
\bigl[ O W K V 
\bigl(1 - O W K V \bigr)
\bigr]^{(I)(\sigma)} . \end{align}
We have assumed throughout, for notational clarity, 
that identical Hubbard projectors are used
for each spin channel, although the 
generalization to $\sigma$-dependent
Hubbard projectors, 
and thus $\sigma$-dependent $O$, $V$, and $W$
matrices, is straightforward.

\subsection{Efficient use of matrix sparsity}

The DFT+$U$ functional of Eqn.~\ref{compactenergy}, 
does not depend on the inter-site occupancy
matrices generated using Hubbard projectors for 
different subspaces, although the generalization
to inter-site occupancies, DFT+$U$+$V$, has been
introduced in Ref.~\onlinecite{0953-8984-22-5-055602}.
In DFT+$U$, these non-local occupancies 
do not contribute to $E_U$ and thus 
should not be computed unnecessarily. 
On the other hand, it is undesirable 
from the point of view of both ease of implementation
and computational efficiency 
to explicitly store separate $V^{(I)}$,
$W^{(I)}$ and $O^{(I)}$ matrices for each site,
thereby necessitating individual matrix products 
for each site before explicit summation in, 
for example, Eqn.~\ref{compactenergy}.

Our solution 
 is to embed these small transformation
matrices into large, though very sparse, $V$, $W$ and
$O$ matrices for the entire system, 
where they then fit seamlessly
into the hierarchical, parallelized, 
sparse algebra 
routines found
in a contemporary 
linear-scaling DFT code~\cite{Hine20091041,hine:114111}. 
The overlap $O$ matrix is block-diagonal
in either its covariant or contravariant form, 
the dimension of each block being
the number of projectors spanning the 
subspace on the site in question, typically $5(7)$ for a
subspace of $3d(4f)$ orbital symmetry.
The Hubbard interaction parameters are also 
placed into a sparse matrix $U$ for the entire system, 
of the same sparsity of $O$, although, in practice, 
diagonal in the simplest
case of a scalar parameter on each site.
The $V$ matrix has the row sparsity of the orbital overlap
matrix $S$, depending on the orbital cutoff radii, 
and the column sparsity of $O$; $W$ is its transpose.

Let us take as an example the computation of
the occupancy matrix given by Eqn.~\ref{Eq:occupmat}.
We henceforth suppress the spin index
for notational simplicity, 
on the understanding that the density kernel, its
derivatives and derivatives with 
respect to it are generally spin-dependent.
Working from left to right,  
temporarily placing a site index before each projector index to 
clarify to which subspace it belongs, 
we first consider the
product \begin{align}
\left(OW\right)_{\quad\;\;\beta}^{(I)m'}={} & \sum_{J}O^{(I)m'
(J)m''}{W}_{(J)m''\beta} \\
\nonumber
={} & O^{(I)m'(I)m''}{W}_{(I)m''\beta},\end{align}
which is a matrix with the same sparsity pattern as $W$ due to the
block-sparsity of the $O$ matrix.
Next, taking the product with the 
density kernel on the right, as per \begin{equation}
\left(OWK\right)^{(I)m'\alpha}=
\left(OW\right)_{\quad\;\;\beta}^{(I)m'}
K^{\beta\alpha},\end{equation}
we see that this matrix has the sparsity of $WK$, dense in the row
index when no density-kernel truncation is applied. 
When kernel truncation is enforced, however, the number of values
which $\alpha$ can take is reduced and the effort needed for the
sum over $\beta$ is diminished.

On the final step, where we compute \begin{equation}
n_{\quad (J)m}^{(I)m'}=\left(OWK\right)^{(I)m'\alpha}
{V}_{\alpha(J)m},\end{equation}
 we accumulate extraneous information 
 on the inter-subspace non-locality
of the density-matrix. Were we to compute this matrix in full and
then consider its square,  
we would find  
that \begin{align}
\sum_{K}n_{\quad (K)m}^{(I)m'}n_{\quad (I)m'}^{(K)m}
\ne n_{\quad(I)m}^{(I)m'}n_{\quad (I)m'}^{(I)m},\end{align}
the former being generated in the full matrix product, while
only the latter is required in Eqn.~\ref{Eq:DFTU}.
This problem is resolved by
always truncating the occupancy matrix \begin{equation}
n_{\quad(I)m}^{(I)m'}=\left(OWK\right)^{(I)m'\alpha}
{V}_{\alpha(I)m}\end{equation}
 to the block-diagonal sparsity pattern 
 as $O$ \emph{in advance} 
 of computing such products, 
 eliminating the off-site occupancies. In practice, the
unnecessary elements are never actually 
computed, and no wasted effort
is incurred, since the sparse algebra system 
computes only elements in the
sparsity pattern of the product matrix~\cite{hine:114111}.

Matrix sparsity thus plays an important role 
in the construction of our linear-scaling
DFT+$U$ method, as it permits DFT+$U$ 
calculations involving a large
number of subspaces to be carried out efficiently. 
We hereafter suppress the site index, both to clarify
the notation and to reflect the fact that the matrix operations
are implemented \emph{in practice} in terms of
calls to sparse algebra subroutines which
take the matrices $V$, $W$, 
$O$ and $U$ as arguments, and not
their site-indexed counterparts.

\section{Optimization of the Density-Matrix}
\label{dm}

In order to minimize the total-energy with respect to the
density-matrix with linear-scaling cost, while affording it the 
variational freedom of a systematically improvable basis,
 it is performed in two nested conjugate gradients minimization loops. 
 We first describe the inner loop, in 
 Section~\ref{densitykernelvariationsimple},
a methodology common to many 
contemporary linear-scaling codes,
where the energy is minimized
with respect to the density kernel for a fixed
set of local orbitals.

In the outer loop, the local orbitals 
$\left\lbrace \ket{ \phi_\alpha } \right\rbrace$ which 
span the Hilbert space available to the density-matrix
are optimized in order to minimize the total-energy.
This technique is used to obviate the choice of a fixed local orbital basis,
and numerous variations have been previously described~\cite{
PhysRevB.66.035119,10.1063/1.1839852,
PhysRevB.67.155108,PhysRevB.69.195113,ozaki:10879,
PhysRevB.47.9973,
PhysRevB.50.4316, PhysRevB.52.1640,PhysRevB.51.1456}.
The orbitals are assumed to be truncated
to some region, in order to allow for linear-scaling cost,
and refined with respect to the underlying basis,
for a fixed density kernel $K^{\alpha\beta}$,
in a manner which is furthermore compatible 
with a linear-scaling method for 
optimizing local orbitals for
unoccupied states recently proposed 
in Ref.~\onlinecite{PhysRevB.84.165131}.
We return to discuss the orbital optimization 
technique in Section~\ref{ngwfvariationsimple}.

Recent success with the DFT+$U$ method  
and its generalization
to inter-site interactions, DFT+$U$+$V$~\cite{0953-8984-22-5-055602}, 
encourages us to think of DFT+$U$ as a true method for first-principles
energetics~\cite{doi:10.1021/jp070549l,
kulik:134314,*kulik:114103,*kulik:094103,PhysRevB.79.125124,PhysRevB.84.115108}. 
We have therefore implemented
the DFT+$U$ forces terms, as well as the total-energy minimization
scheme, in the \textsc{ONETEP} code of which
the capability of accurately optimizing geometries 
 has been previously demonstrated~\cite{PhysRevB.83.195102}.
We describe the required methodology in Appendix~\ref{ionicvariations}.

\subsection{Kernel Optimization}
\label{densitykernelvariationsimple}

Minimization of the energy with respect to the density kernel 
is typically carried out, in practice, using a generalization of the 
Li-Nunes-Vanderbilt (LNV) technique~\cite{PhysRevB.47.10891,
*PhysRevB.50.17611,
*PhysRevB.47.10895}, 
which simultaneously 
drives the density-matrix to idempotency while 
it evolves towards commutativity
with its corresponding Kohn-Sham Hamiltonian.
In this section however, for clarity, we assume that the 
energy may be straightforwardly minimized
with respect to the density kernel. 
We return to the adaptations to the
density kernel optimization method required for
 idempotency preservation
in Appendix~\ref{densitykernelvariations}. 

The DFT+$U$ contribution to the Hamiltonian is thus simply given
by the derivative of the DFT+$U$ energy term of Eqn.~\ref{Eq:DFTU}
with respect to an arbitrary density kernel, that is \begin{align}
H_{\beta\alpha}^{U} &{}=
\frac{U}{2}\Big\lbrace\frac{\partial n_{\;\; m}^{m}}{\partial K^{\alpha\beta}}
-\frac{\partial n_{\;\; m}^{m'}}{\partial K^{\alpha\beta}}n_{\;\; m'}^{m}-n_{\;\; m}^{m'}\frac{\partial n_{\;\; m'}^{m}}{\partial K^{\alpha\beta}}\Big\rbrace. 
\end{align}
 In order to simplify this derivative, we begin by noting
that the partial derivative of the occupation 
matrix with respect to the density kernel is  
given by \begin{align}
\frac{\partial{n_{\;\; m}^{m'}}}{\partial K^{\alpha\beta}} & =\frac{\partial}{\partial K^{\alpha\beta}}\left[O^{m'm''}W_{m''\gamma}^{}K^{\gamma\delta}V_{\delta m}^{}\right]
\\ & 
=O^{m'm''}W_{m''\gamma}^{}\delta_{\alpha}^{\gamma}\delta_{\beta}^{\delta}V_{\delta m}^{}\nonumber \\
 & =O^{m'm''}W_{m''\alpha}^{}V_{\beta m}^{}.\nonumber \end{align}
 The trace of this derivative over Hubbard projectors 
 gives the covariant, local-orbital representation of the sum
 of projections over subspaces, which is a Hermitian tensor
by construction, given by \begin{align}
\frac{\partial{n_{\;\; m}^{m}}}{\partial K^{\alpha\beta}} & =V_{\beta m}^{}O^{mm''}W_{m''\alpha}^{}\equiv P_{\beta\alpha}^{}.
\label{Eq:myproj}\end{align}
 It follows that the products of the occupancy matrix and its derivative,
each always computed in terms of the Hubbard projector
indices, in practice, since
there they have the block-diagonal sparsity pattern of $O$,
are given by \begin{align}
\frac{\partial{n_{\;\; m''}^{m'}}}{\partial K^{\alpha\beta}}n_{\;\; m}^{m''} & {}=\left(OW\right)_{\;\;\alpha}^{m'}\left(PKV\right)_{\beta m}^{}\;\;\mbox{and}\\
n_{\;\; m''}^{m'}\frac{\partial{n_{\;\; m}^{m''}}}{\partial K^{\alpha\beta}} & {}=
\left(OWKP\right)_{\;\; \alpha}^{m'} V_{\beta m'}^{}. \end{align}
As a result, the DFT+$U$ term in the covariant Hamiltonian, denoted by $H^{U}$, 
may be succinctly expressed as 
\begin{equation}
H_{\beta\alpha}^{U}=\frac{\partial E_{U}^{}}{\partial K^{\alpha\beta}}=\frac{U^{}}{2}\left(P-2PKP\right)_{\beta\alpha}^{}.\label{Eq:mylittleham}\end{equation}
 The DFT+$U$ contribution to the total-energy is
efficiently computed, correspondingly, using the trace \begin{equation}
E_{U}^{}=\frac{U^{}}{2}\left(PK-PKPK\right)_{\alpha}^{\;\;\alpha}.
\label{Eq:mylittleenergy}\end{equation}

The DFT+$U$ Hamiltonian and total-energy terms
are added to their
uncorrected DFT counterparts, giving 
$H_{\alpha \beta} = H_{\alpha \beta}^{DFT} + H_{\alpha \beta}^{U}$
and $E = E_{DFT} + E_U$, 
respectively, and similarly for the
independent-particle, or ``band-structure'' energy
$E^{IP}=E_{DFT}^{IP}+E_{U}^{IP}$, where
$E_{DFT} \ne E_{DFT}^{IP}
= H^{DFT}_{\alpha \beta} K^{\beta\alpha}$ and
$E_{U}  \ne E_{U}^{IP} = 
H^U_{\alpha\beta} K^{\beta\alpha}$.

 For a refinement of the auxiliary density kernel $K^{\alpha\beta}$,
any update to it must also be a contravariantly transforming tensor, as
noted in Refs.~\onlinecite{Artacho,White1997133}. In order to
provide such a search direction, it is necessary that we pre- and
post-multiply the covariant gradient of Eqn.~\ref{Eq:mylittleham}
with the contravariant
metric tensor on the orbitals, that is their inverse overlap matrix
evaluated at the point at which the gradient itself is computed, to
give \begin{align}
G^{\alpha\beta} & {}=\left(S^{-1}\right)^{\alpha\gamma}
H_{\gamma\delta}\left(S^{-1}\right)^{\delta\beta}.
\label{Eq:mygradsimple}\end{align}
The inner product of two second-order tensors,
$X$ and $Y $, 
is defined with respect to the metric $S$ on the local orbitals, so that
\begin{align}
\langle X \rvert Y \rangle_S 
=  X^{\alpha \beta} Y_{\beta \alpha }
=  X^{\alpha \beta} S_{\beta \gamma } Y^{\gamma \delta }
S_{\delta \alpha } .
\end{align}
This allows us to define the search direction norm
$\lVert G \rVert_S = \langle G \rvert G \rangle^{1/2}_S$, and the 
conjugacy condition $\langle G_{i+1} \rvert G_i \rangle_S = 0$, 
and thus to  minimize the total-energy
by iteratively updating the density kernel according to
\begin{align}
 K^{\alpha \beta}_{i+1} \rightarrow
K^{\alpha \beta}_{i} +  \lambda_{i+1} G^{\alpha \beta}_{i+1},
\end{align}
where the optimal step lengths 
$\{  \lambda_{i} \} $ are computed using an appropriate 
non-linear conjugate gradients algorithm.

\subsection{Orbital optimization}

\label{ngwfvariationsimple}

We now consider the optimization of the local orbitals, specifically
the DFT+$U$ contribution to total-energy variations with respect
to the expansion coefficients of the orbitals in the underlying variational
basis. We again assume that the energy may be
directly minimized in this section, for simplicity, 
returning to the alterations necessary for idempotency 
preservation under local-orbital optimization
in Appendix~\ref{ngwfvariations}.
This procedure occurs in the outer of the two energy minimization loops in
the \textsc{ONETEP} code used here, however the results of this section
apply to any technique which 
optimizes its representation functions for minimal energy, 
such as those described in Refs.~\onlinecite{
PhysRevB.66.035119,10.1063/1.1839852,
PhysRevB.67.155108,PhysRevB.69.195113,ozaki:10879,
PhysRevB.47.9973,
PhysRevB.50.4316, PhysRevB.52.1640,PhysRevB.51.1456}.

It is clear from Eqn.~\ref{Eq:mylittleenergy} that the 
derivative of the total-energy with respect to the 
expansion of the local orbitals on the grid
(or, in general, the basis) may explicitly depend only on
the matrix elements of the projection $P$, 
defined in Eqn.~\ref{Eq:myproj}, so that
\begin{align}
\frac{\partial E_{U}}{\partial\phi_{\alpha}\left(\mathbf{r}\right)} & {}=
\frac{\partial E_{U}}{\partial P_{\beta\gamma}^{}}\frac{\partial 
P_{\beta\gamma}^{}}{\partial\phi_{\alpha}\left(\mathbf{r}\right)}.
\label{Eq:NGWFGRADsimple}\end{align}
Since this derivative involves the expansion of the Hubbard
projections on the grid, it incurs changes beyond simple linear mixing
of the orbitals. Evaluating this, 
we first take the action of the DFT+$U$ Hamiltonian  contribution
on the subspace projections, that is
\begin{align}  \label{Eq:PROJGRAD}
\frac{\partial E_{U}}{\partial P_{\beta\gamma}^{}} & =\frac{U^{}}{2}\frac{\partial}{\partial P_{\beta\gamma}^{}}\left[\left(P^{}K-P^{}KP^{}K\right)_{\alpha}^{\;\;\alpha}\right]
 \\ \nonumber
 & =K^{\gamma\delta}H_{\delta\epsilon}^{U}P^{\epsilon\beta}; \quad
P^{\alpha \beta} = \left( P^{-1} \right)^{\alpha \beta}  .
\end{align}
The Hubbard projection operators depend
explicitly on the covariant orbitals which overlap with their corresponding
Hubbard projectors (and Hubbard projector duals) and this dependence
may be expressed as \begin{align} 
\frac{\partial P_{\beta\gamma}^{}}{\partial\phi_{\alpha}\left(\mathbf{r}\right)} & =\frac{\partial}{\partial\phi_{\alpha}\left(\mathbf{r}\right)}\left[V_{\beta m}^{}O^{mm'}W_{m'\gamma}^{}\right]\\
 & =\delta_{\beta}^{\alpha}\varphi_{m}^{}\left(\mathbf{r}\right)O^{mm'}W_{m'\gamma}+V_{\beta m}^{}O^{mm'}\varphi_{m'}^{}\left(\mathbf{r}\right)\delta_{\gamma}^{\alpha}.\nonumber 
 \end{align}
Combining this result with Eqs.~\ref{Eq:NGWFGRADsimple} and~\ref{Eq:PROJGRAD}, 
we may compute the DFT+$U$
term in the local-orbital gradient,
\begin{align} \label{Eq:gridgradient}
\frac{\partial E_{U}}{\partial\phi_{\alpha}\left(\mathbf{r}\right)} & {}=2 K^{\alpha\delta}H_{\delta\epsilon}^{U}P^{\epsilon\beta}V_{\beta m}^{}O^{mm'}\varphi_{m'}^{}\left(\mathbf{r}\right)
\\ \nonumber 
 & {}=2 K^{\alpha\delta}
 V_{\delta m''}  H^{m'' m'}
 \varphi_{m'}^{}\left(\mathbf{r}\right),
\end{align}
 where formally, though never explicitly in practice,
\begin{align}
 H_U^{m m'} = O^{m m''} W_{m'' \alpha} S^{\alpha \beta } H^U_{\beta \gamma}
S^{\gamma \delta} V_{\delta m'''} O^{m''' m'} .
 \end{align}
Due to the subspace-localized nature of the DFT+$U$
correction in the tensorial 
representation~\cite{PhysRevB.83.245124}, 
only those local orbitals $\lvert \phi_{\delta} \rangle$ in
Eqn.~\ref{Eq:gridgradient} which explicitly overlap
with the Hubbard projectors 
$\lvert \varphi_{m'}^{}\rangle$ contribute and 
thus require summation over.

Since, crucially, we require a covariantly transforming orbital update
in order to improve upon those functions, to preserve their tensorial
character, the above contravariant gradient must be multiplied 
with the covariant metric tensor in order to provide the necessary
covariant DFT+$U$ orbital search direction term, 
given by \begin{align}
g^U_{\alpha}\left(\mathbf{r}\right)=2S_{\alpha\beta}
K^{\beta\delta}
 V_{\delta m''}  H^{m'' m'}
 \varphi_{m'}^{}\left(\mathbf{r}\right).
\label{Eq:uncorrectedgrad} \end{align}
This contribution may then be combined with 
the uncorrected DFT search direction,
giving the total $ \ket{g_{\alpha }} = \ket{ g^{DFT}_{\alpha} }
+ \ket{ g^U_{\alpha} }$.
The inner product and norm of first-order tensors, or orbitals,
$\ket{ x}$ and $\ket{y} $, 
are defined such that
\begin{align}
\langle x \rvert y \rangle_S 
&{}=  \langle x^{\alpha } \rvert y_{\alpha } \rangle
=  \langle x_{\alpha } \rvert S^{\alpha \beta  } \lvert y_{\beta } \rangle; \\
\lVert x \rVert_S &{}= \langle x \rvert x \rangle^{1/2}_S,
\end{align}
and, using these definitions, the total-energy may be minimized
by iteratively updating the orbitals,
where the $\{  \mu^{i} \} $ are computed using one of many
available non-linear conjugate gradients algorithms,  according to
\begin{align}
\ket{ \phi_{\alpha}^{i+1} } \rightarrow
\ket{ \phi_{\alpha }^{i} } +  \mu^{i+1} \ket{ g_{\alpha }^{i+1} }.
\end{align}
At each such orbital update step, in practice, 
we carry out a complete re-optimization of
the density kernel, according to the procedure of
Sub-section~\ref{densitykernelvariationsimple}.

\section{Application to nickel oxide nano-clusters}
\label{nickeloxidetests}

We performed scaling tests on NiO nano-clusters of varying size, comparing
the computational effort required for DFT+$U$ and uncorrected DFT
calculations. An antiferromagnetic insulator, NiO is a well known example
where LSDA-type 
approximations~\cite{PhysRevB.23.5048}  fail to qualitatively
reproduce the correct insulating gap and local magnetic moments 
(of between $1.64~\mu_{B}$ and 
$1.9~\mu_{B}$~\cite{PhysRevB.71.035105})
due to a poor
description of $3d$ orbital localization. The gap, of approximately
$4~$eV, is of predominantly
Mott-Hubbard type since it persists above the 
N\'{e}el temperature~\cite{PhysRevB.50.5041}, albeit
with a significant charge-transfer component~\cite{PhysRevLett.53.2339}. 
It is thus successfully  
recovered by a number of methods, which either include many-body Coulomb 
correlation effects explicitly, such as 
LDA+DMFT~\cite{PhysRevB.74.195114}, 
or introduce an appropriate derivative discontinuity
with respect to occupancy at the single-particle level, examples including
unrestricted Hartree-Fock~\cite{PhysRevB.50.5041} and
the self-interaction corrected local 
density approximation~\cite{PhysRevLett.65.1148}. 
The correct description of the physics of NiO 
was an early success for DFT+$U$, the method of interest here,
and this has been repeated using numerous 
functional forms~\cite{PhysRevB.44.943,*PhysRevB.48.16929,
PhysRevB.57.1505,PhysRevB.62.16392,
PhysRevB.58.1201,PhysRevB.71.035105,0953-8984-22-5-055602}.

The method described in this article 
has previously been successfully  
applied to bulk NiO~\cite{PhysRevB.83.245124}.
For a demonstration of computational scaling, 
we have chosen spherical nano-clusters of NiO with
even numbers of nickel ions, so that an open-shell singlet multiplicity,
analogous to the bulk antiferromagnetic ground state, could be tentatively
assumed. We may expect that a transition to a ferrimagnetic or ferromagnetic
state occurs below some critical cluster size, as it has been
predicted for very small iron oxide clusters of interest for data-storage
technology~\cite{PhysRevB.80.085107,PhysRevB.81.075403}. 

\begin{figure}
\includegraphics[width=9.0cm]{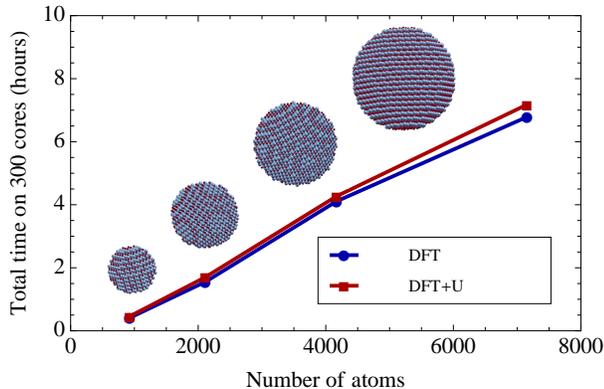} 
\caption{(Color online) Scaling of the energy minimization algorithm
for NiO nanoclusters of increasing size.
Timings are for three density kernel 
optimization steps and one orbital
optimization step, comparing DFT and DFT+$U$ calculations. 
Simulations were performed on $300$ Intel
Westmere $2.67$~Ghz cores connected using quad data 
rate Infiniband.}
\label{Fig:NiO_Scaling_illustrated.eps} 
\end{figure}

Run-time parameters
included a $500~$eV equivalent plane-wave cutoff energy, a spin polarized
density kernel cutoff at $25~$a$_{0}$, the LSDA exchange-correlation
functional~\cite{PhysRevB.23.5048}, nine local orbitals (NGWFs) 
for each nickel
ion and four each for oxygen, all with $7.5~$a$_{0}$ cutoff radii,
and norm-conserving pseudopotentials~\cite{opium}.
Atomic Hubbard projectors of hydrogenic form were used. 
Since calculations
on nano-clusters of varying sizes are expected to exhibit differing
convergence behavior, the energy minimization algorithm was simply
run for a fixed number of iterations. 
One orbital optimization
step and three density kernel steps, with three penalty-functional
idempotency corrections iterations at each of the latter, were allowed.
Orbital overlap matrix inversion was carried out using a sparse matrix
implementation of Hotelling's algorithm~\cite{PhysRevB.64.195110}
and a cubic supercell of length three times the diameter of each nano-cluster
was used, up to a maximum supercell length of 
approximately $300~$a$_{0}$.

\subsection{Scaling of computational effort for DFT+$U$}

Algorithmic timing data for \textsc{ONETEP} energy minimization of
NiO nano-clusters, containing up to $7,153$ atoms across $300$ Intel
Westmere $2.67$~Ghz cores, is shown in Fig.~\ref{Fig:NiO_Scaling_illustrated.eps}.
A reasonable linear fit was obtained for the timing; with a slightly
negative fitted time intercept at $450-500$ atoms indicating a very efficient
initialization of the pre-requisite data in these calculations. The
NiO nano-clusters in question do not represent a favorable case for the DFT+$U$ method, since approximately half of the ions host correlated
subspaces. Nonetheless, we observed a very small increase in computational
time when the DFT+$U$ functionality was invoked, at approximately
$5-6\%$, and preservation of linear-scaling performance.

\begin{figure}
\includegraphics[width=8.5cm]{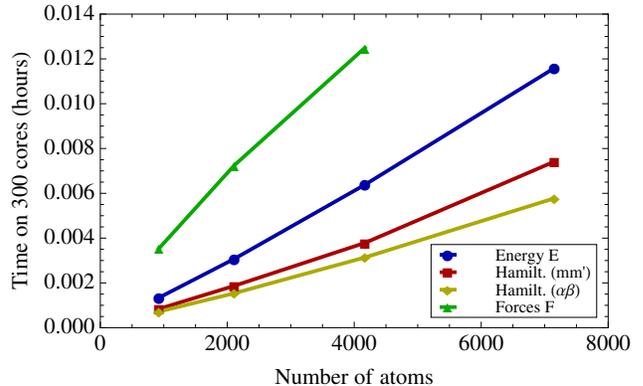} \caption{(Color 
online) Computational time spent in subroutines
associated with the DFT+$U$ functionality
in the tests shown in
Fig.~\ref{Fig:NiO_Scaling_illustrated.eps}.
Specifically timings shown are
for computing 
the DFT+$U$ energy of Eqn.~\ref{Eq:DFTU}, 
the Hamiltonian matrix of Eqn.~\ref{Eq:mylittleham} in
its Hubbard projector and 
local orbital representations,
and the ionic forces given by Eqn.~\ref{Eq:DFTUforce}.}
\label{Fig:NiO_Hubbard.eps} 
\end{figure}

Timings for generating the DFT+$U$ Hamiltonian
and its contribution to the total-energy and forces,
for those calculations
which fell within memory resources, are depicted
in Fig.~\ref{Fig:NiO_Hubbard.eps}. This indicates that no direct
DFT+$U$ functionality appreciably deviates from linear-scaling behavior.
We note, in particular,
that the total time spent in these 
DFT+$U$ specific subroutines makes
up only a small fraction of the increase in cost incurred by DFT+$U$,
at less than $1\%$ of the total computational time.

\begin{figure}
\includegraphics[width=8.5cm]{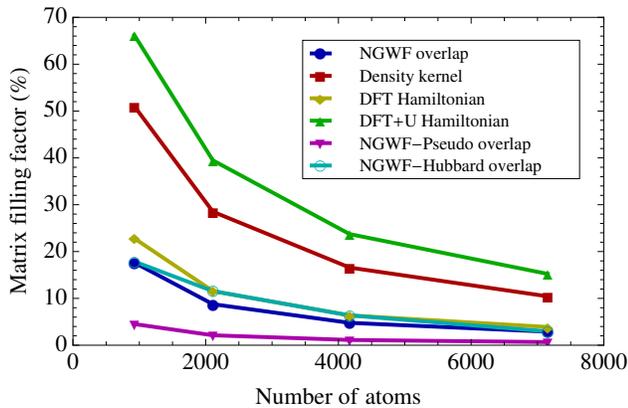} \caption{(Color online) Nano-cluster size dependence of filling factors of
principal matrices affecting the computational cost of linear-scaling
DFT and DFT+$U$. These, namely, are the overlap between orbitals, the
density kernel, the Hamiltonian matrices conventional for DFT and
for DFT+$U$, and the overlap matrices between the local orbitals
and the non-local pseudopotential projectors and Hubbard projectors.}
\label{Fig:NiO_Filling.eps} 
\end{figure}

In order to understand where the dominant contribution to the 
DFT+$U$ cost originates, 
since it is not directly in the DFT+$U$ subroutines themselves, we
direct the reader to Fig.~\ref{Fig:NiO_Filling.eps}, where the system
dependent sparse matrix filling 
is quantified. In a conventional DFT calculation, the sparsity
of the Hamiltonian matrix is dominated by the orbital representation
of the non-local pseudopotential, proportional to the product of the
overlap matrix between the orbitals and the non-local pseudopotential
projectors with its transpose. In essence, pairs of orbitals which
overlap with a common non-local projector contribute to 
the energy, and the same holds for the 
Hubbard projectors of DFT+$U$. While
non-local pseudopotential projectors tend to have cutoff radii not
in excess of approximately $2~$a$_{0}$, Hubbard projectors of $3d$
symmetry may require greater cutoff radii 
    (for hydrogenic orbitals of effective nuclear 
charge in the typical range for transition-metals,
$Z=\left\lbrace 8,9,10,11\right\rbrace$, the 
normalization
spillage at $2~$a$_{0}$
is $\left\lbrace 9.3,4.6,2.1,0.9\right\rbrace $\%). 
In our calculations the projector cutoff radius is set
equal to that of the local orbitals, $7.5~$a$_{0}$.

\begin{figure}
\includegraphics[width=8.5cm]{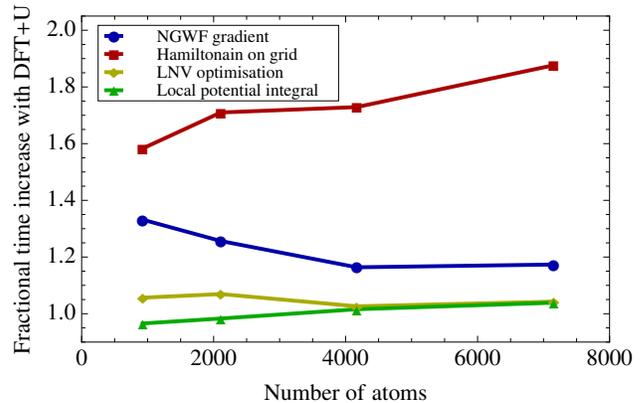} \caption{(Color 
online) Fractional change in time expended on energy-minimization 
operations when the DFT+$U$ 
functionality is activated. Namely, these are calculation of 
the local orbital (NGWF) gradient, expansion of the Hamiltonian
in the basis, kernel optimization using the LNV method, and 
calculation of the matrix elements of the local potential.}
\label{Fig:NiO_Non_Hubbard.eps} 
\end{figure}

This increased Hamiltonian filling
 has consequences additionally for the calculation of
energy gradients, as indicated in Fig.~\ref{Fig:NiO_Non_Hubbard.eps},
which shows the fractional change in time spent in 
carrying out certain energy minimization operations.
Most notably, it takes close to twice as 
much effort to calculate its expansion
on the \emph{psinc} grid due to the 
inclusion of Hubbard projectors in the Hamiltonian.
The dominant part of the overall expense of the calculations
is from operations on large matrices, however, so that grid
expansion of the Hamiltonian is not significant for large systems.
The incurred increase in the filling of the Hamiltonian matrix in
DFT+$U$ over DFT, and also in the expense of computing its products
with quantities such as the density kernel and its expansion on the
underlying, systematic plane-wave basis, is thus largely responsible for
the observed, albeit moderate, increase 
in computational expense indirectly introduced
by DFT+$U$, from $1\%$ to $5-6\%$.

\section{Concluding remarks}

We have detailed a linear-scaling implementation of the DFT+$U$ method
for treating strongly-correlated systems from first-principles. The
formalism is generally appropriate to methods which minimize the energy
with respect to the single-particle density-matrix, and allows for
the optimization of both nonorthogonal Hubbard 
projectors~\cite{PhysRevB.82.081102} and ionic positions.

The preservation of linear-scaling performance
on metal-oxide nano-clusters in excess of $7,000$ atoms is demonstrated.
For systems of this type, with a high density of correlated
sites, the increase in computational pre-factor remains rather modest.
The DFT+$U$ functionality, furthermore, incurs negligible cost
in large systems comprising only a small number of Hubbard subspaces.

Ground state calculations employing our method
have previously been demonstrated 
on both bulk and molecular strongly interacting
systems~\cite{PhysRevB.83.245124,PhysRevB.82.081102},
with further examples on large-scale systems
such as dilute magnetic semiconductor 
(Ga,Mn)As~\cite{forthcomingGaMnAs}
and disordered VO$_2$~\cite{forthcomingVO2}, 
using an extension of the method to DFT+DMFT, forthcoming.
Further examples of candidate
systems include organometallic 
molecules, such as metalloproteins and
molecular magnets, where the method is
particularly efficient for a low density of
strongly interacting subspaces, and solids such as
magnetic heterostructures, 
defective and doped oxides or 
catalytic interfaces with oxide surfaces. 
We envisage that the technique described 
may aid in bringing linear-scaling DFT to bear on
more challenging systems than those 
to which it is has been typically applied to date.

\begin{acknowledgments}
All figures are reproduced with 
permission from Ref.~\onlinecite{thesis}.
We are grateful to Emilio Artacho and Adrian 
Ionescu for helpful discussions.
D.D.O'R acknowledges the support of
EPSRC and the National University of Ireland. 
N.D.M.H and A.A.M. acknowledge 
the support of EPSRC (Grant EP/G05567X/1), 
and A.A.M further acknowledges support from RCUK.
M.C.P. acknowledges the support of
EPSRC (Grants EP/G055904/1, EP/F032773/1).
Calculations were performed on
the Cambridge HPCS Darwin computer.
\end{acknowledgments}

\appendix

\section{Preservation of density-matrix purity under kernel optimization}

\label{densitykernelvariations}

In the LNV method~\cite{PhysRevB.47.10891,*PhysRevB.50.17611,*PhysRevB.47.10895},
the Kohn-Sham density kernel is related to the auxiliary density kernel
via one iteration of the McWeeny purification transform, that is \begin{equation}
K^{\alpha\beta}=\left(3LSL-2LSLSL\right)^{\alpha\beta}.\label{Eq:LNVorig}\end{equation}
 In our treatment of DFT+$U$, we go a step further and provide the
more general expressions needed for the HSMP~\cite{0953-8984-20-29-294207}
variant of the LNV method, in which the density kernel $\tilde{K}$
is expressed as a purified and normalized auxiliary density kernel,
explicitly \begin{align}
\tilde{K}^{\alpha\beta} &{}= \frac{N}{S_{\gamma\delta}K^{\delta\gamma}}K^{\alpha\beta}\\
&{}= \frac{N \left(3LSL-2LSLSL\right)^{\alpha\beta}}{S_{\gamma\delta}\left(3LSL-2LSLSL\right)^{\delta\gamma}}, \nonumber \end{align}
 where $N$ is the correct occupancy for the spin channel in question.
The kernel renormalization introduces terms in the gradient proportional
to an effective chemical potential, projecting out any first-order
changes to the electron number, driving the density kernel $\tilde{K}$
towards both normalization and idempotency as the energy is minimized.

To locate the derivative of the DFT+$U$ energy term with respect
to the auxiliary density kernel, stressing that it is
computed strictly using the purified and renormalized density kernel,
we make use of the chain-rule for matrix derivatives to write
\begin{equation}
\frac{\partial E_{U}}{\partial L^{\alpha\beta}}=\frac{\partial E_{U}}{\partial K^{\gamma\delta}}\frac{\partial K^{\gamma\delta}}{\partial L^{\alpha\beta}},\end{equation}
 where we carefully note that the partial derivative with respect
to a doubly contravariant tensor is a doubly covariant tensor with
indices once permuted to allow for the complex case. It may be readily
shown, using Eqn.~\ref{Eq:LNVorig}, that the latter term is given
by \begin{align}
\frac{\partial K^{\gamma\delta}}{\partial L^{\alpha\beta}} & {}=3\left(\delta_{\alpha}^{\gamma}S_{\beta\epsilon}L^{\epsilon\delta}+L^{\gamma\epsilon}S_{\epsilon\alpha}\delta_{\beta}^{\delta}\right)\label{Eq:readily}\\
 & \quad-2\left(\begin{array}{cc}
\delta_{\alpha}^{\gamma}S_{\beta\epsilon}L^{\epsilon\zeta}S_{\zeta\eta}L^{\eta\delta}\\
+L^{\gamma\epsilon}S_{\epsilon\alpha}S_{\beta\zeta}L^{\zeta\delta}+L^{\gamma\epsilon}S_{\epsilon\zeta}L^{\zeta\eta}S_{\eta\alpha}\delta_{\beta}^{\delta}\end{array}\right).\nonumber \end{align}

The derivative of the DFT+$U$ energy with respect to the purified
density kernel $K$ 
may be broken into products of derivatives, and rearranged as follows
\begin{align}
\frac{\partial E_{U}}{\partial K^{\gamma\delta}} & {}=\frac{\partial}{\partial\tilde{K}^{\zeta\epsilon}}\left[E_{U}\left(\tilde{K}\right)\right]\frac{\partial\tilde{K}^{\zeta\epsilon}}{\partial K^{\gamma\delta}}\\
 & {}=H_{\epsilon\zeta}^{U}\frac{N}{\left(S_{\eta\theta}K^{\theta\eta}\right)}\left\lbrace \delta_{\gamma}^{\zeta}\delta_{\delta}^{\epsilon}-\frac{K^{\zeta\epsilon}}{\left(S_{\iota\kappa}K^{\kappa\iota}\right)}S_{\delta\gamma}\right\rbrace .\nonumber \end{align}
We may next write the gradient with respect to the density kernel
in terms of a preconditioned contribution to the Hamiltonian, denoted by
\begin{align}
\tilde{H}_{\delta\gamma}^{U}=H_{\delta\gamma}^{U}-\mu^{U}S_{\delta\gamma};
\quad
\mu^{U}=\frac{H_{\epsilon\zeta}^{U}K^{\zeta\epsilon}}{\left(S_{\iota\kappa}K^{\kappa\iota}\right)},
\label{Eq:precon}\end{align} 
where $\mu^{U}$ is identified as the DFT+$U$ correction to the chemical potential,
 since \begin{align}
\frac{\partial E_{U}}{\partial K^{\gamma\delta}} & {}=\frac{N}{\left(S_{\eta\theta}K^{\theta\eta}\right)}\left\lbrace H_{\delta\gamma}^{U}-\frac{H_{\epsilon\zeta}^{U}K^{\zeta\epsilon}}{\left(S_{\iota\kappa}K^{\kappa\iota}\right)}S_{\delta\gamma}\right\rbrace  \\ \nonumber
 & {}=\frac{N}{\left(S_{\eta\theta}K^{\theta\eta}\right)}\tilde{H}_{\delta\gamma}^{U}.\end{align}
It is worth noting that, just as the DFT+$U$ independent-particle
energy correction, $E_{U}^{IP}=H_{\alpha\beta}^{U}\tilde{K}^{\beta\alpha}$,
does not equal the energy term $E_{U}$, so the energy correction
entering into the computation of $\mu^{U}$ is not identical to the
DFT+$U$ correction to the total-energy. 

The required DFT+$U$  energy gradient is provided
by the product of the preconditioned term in the Hamiltonian and the
derivative of the density kernel with respect to its auxiliary counterpart,
given by \begin{align}
\frac{\partial E_{U}}{\partial L^{\alpha\beta}} & {}=\frac{N}{\left(S_{\iota\kappa}K^{\kappa\iota}\right)}\tilde{H}_{\delta\gamma}^{U}\frac{\partial K^{\gamma\delta}}{\partial L^{\alpha\beta}}.\label{Eq:mygrad1}\end{align}
Finally, combining Eqs.~\ref{Eq:readily} and ~\ref{Eq:mygrad1}, 
and recalling Eqn.~\ref{Eq:mygradsimple}, we arrive
at the DFT+$U$ contribution to the contravariant density kernel gradient,
\begin{align}
G_{U}^{\alpha\beta} & {}=\left(S^{-1}\right)^{\alpha\gamma}\frac{\partial E_{U}}{\partial L^{\delta\gamma}}\left(S^{-1}\right)^{\delta\beta} \\ \nonumber
 & {}=\frac{N}{S_{\gamma\delta}\left(3LSL-2LSLSL\right)^{\delta\gamma}}\\
 & \quad\times \left\lbrace \begin{array}{ccc}
3\left(S^{-1}\tilde{H}L+L\tilde{H}S^{-1}\right)\\
-2L\tilde{H}L\\
-2\left(S^{-1}\tilde{H}LSL+LSL\tilde{H}S^{-1}\right) \end{array}\right\rbrace^{\alpha\beta}.\nonumber \end{align}

\section{Preservation of density-matrix purity under orbital optimization}

\label{ngwfvariations}

As in the case of the density kernel gradient, the orbital gradient
is calculated using the purified and renormalized density kernel, and
so it contains a preconditioning term which drives the trace of the density-matrix
to the correct occupancy of the system. 
The energy derivative with respect to covariant orbitals may be decomposed
 as \begin{align}
\frac{\partial E_{U}}{\partial\phi_{\alpha}\left(\mathbf{r}\right)} 
&{}=\frac{\partial E_{U}}{\partial\tilde{K}^{\beta\gamma}}\left(\frac{\partial\tilde{K}^{\beta\gamma}}{\partial K^{\delta\epsilon}}\frac{\partial K^{\delta\epsilon}}{\partial S_{\zeta\eta}}+\frac{\partial\tilde{K}^{\beta\gamma}}{\partial S_{\zeta\eta}}\right)\frac{\partial S_{\zeta\eta}}{\partial\phi_{\alpha}\left(\mathbf{r}\right)} \nonumber \\
 & \qquad+\frac{\partial E_{U}}{\partial P_{\beta\gamma}^{}}\frac{\partial P_{\beta\gamma}^{}}{\partial\phi_{\alpha}\left(\mathbf{r}\right)}.\label{Eq:NGWFGRAD}\end{align}

The terms contained in parentheses in 
Eqn.~\ref{Eq:NGWFGRAD}
may be evaluated, respectively yielding \begin{align}
\label{Eq:threeterms}
\frac{\partial\tilde{K}^{\beta\gamma}}{\partial S_{\delta\epsilon}} & {}=\frac{N}{\left(S_{\theta\iota}K^{\iota\theta}\right)}\left\lbrace -\frac{K^{\beta\gamma}}{\left(S_{\kappa\lambda}K^{\lambda\kappa}\right)}K^{\epsilon\delta}\right\rbrace ,\\
\frac{\partial\tilde{K}^{\beta\gamma}}{\partial K^{\delta\epsilon}} & {}=\frac{N}{\left(S_{\theta\iota}K^{\iota\theta}\right)}\left\lbrace \delta_{\delta}^{\beta}\delta_{\epsilon}^{\gamma}-\frac{K^{\beta\gamma}}{\left(S_{\kappa\lambda}K^{\lambda\kappa}\right)}S_{\epsilon\delta}\right\rbrace, \quad\mbox{and} \nonumber \\
\frac{\partial K^{\delta\epsilon}}{\partial S_{\zeta\eta}} & {}=3L^{\delta\zeta}L^{\eta\epsilon}-2L^{\delta\zeta}\left(LSL\right)^{\eta\epsilon}-2\left(LSL\right)^{\delta\zeta}L^{\eta\epsilon}.\nonumber
 \end{align}
The covariant metric explicitly
depends only on the covariant orbitals, so that
\begin{equation}
\frac{\partial S_{\zeta\eta}}{\partial\phi_{\alpha}\left(\mathbf{r}\right)}=
\delta_{\zeta}^{\alpha}\phi_{\eta}\left(\mathbf{r}\right)+\delta_{\eta}^{\alpha}
\phi_{\zeta}\left(\mathbf{r}\right).\end{equation}

Contraction of the DFT+$U$ term in the Hamiltonian and the terms
in Eqn.~\ref{Eq:threeterms} provides a tensor 
$\tilde{Q}$ which represents a 
contribution to the local orbital gradient purely due to mixing
among the orbitals, given by \begin{align}
\tilde{Q}^{\eta\zeta} & {}=H_{\gamma\beta}^{U}\left(\frac{\partial\tilde{K}^{\beta\gamma}}{\partial K^{\delta\epsilon}}\frac{\partial K^{\delta\epsilon}}{\partial S_{\zeta\eta}}+\frac{\partial\tilde{K}^{\beta\gamma}}{\partial S_{\zeta\eta}}\right)\label{Eq:myQ}\\
 & {}=\frac{N}{\left(S_{\theta\iota}K^{\iota\theta}\right)}\left\lbrace \begin{array}{cc}
3L\tilde{H}L-2L\tilde{H}LSL-2LSL\tilde{H}L\\
-\mu^{U}K\end{array}\right\rbrace ^{\eta\zeta}.\nonumber \end{align}

To conclude, by combining Eqs.~\ref{Eq:gridgradient},
~\ref{Eq:NGWFGRAD}, and~\ref{Eq:myQ},
the contravariant gradient of the DFT+$U$ energy with respect to
the orbitals is given by 
\begin{align}
\frac{\partial E_{U}}{\partial\phi_{\alpha}\left(\mathbf{r}\right)}=2\left(\tilde{K}^{\alpha\zeta}
 V_{\zeta m''}  \tilde{H}^{m'' m'}
 \varphi_{m'}^{}
 +
 \tilde{Q}^{\alpha\zeta}\phi_{\zeta}
 \right)\left(\mathbf{r}\right),\end{align}
which is then transformed to the required covariant form, in the same
manner as per Eq.~\ref{Eq:uncorrectedgrad}, giving
\begin{align}
g^U_{\alpha}\left(\mathbf{r}\right) ={}&
 2S_{\alpha\beta} \tilde{K}^{\beta\zeta}
 V_{\zeta m''}  \tilde{H}^{m'' m'}
\varphi_{m'}^{} \left(\mathbf{r}\right) 
\\ {}&
+ 2S_{\alpha\beta} \tilde{Q}^{\beta\zeta}\phi_{\zeta}
\left(\mathbf{r}\right). \nonumber
 \end{align}

\section{Ionic forces}

\label{ionicvariations}

We may assume that the ground-state 
density is located for a given ionic configuration
 before the forces are computed, so that the total-energy
is variationally minimized with respect to both the orbital expansion
coefficients and the matrix elements of the density kernel. The DFT+$U$
correction then contributes to the ionic forces only via the spatial
dependence of the Hubbard projection operators, that is for the ion
labeled $j$, \begin{align}
\mathbf{F}_{U}^{j}=-\frac{\partial E_{U}}{\partial\mathbf{R}_{j}}=-\frac{\partial E_{U}}{\partial P_{\alpha\beta}^{}}\frac{\partial P_{\alpha\beta}^{}}{\partial\mathbf{R}_{j}}.\label{Eq:FORCE}\end{align}
The lattermost derivative may be expressed in terms 
of gradients of the covariant projectors and contravariant
subspace metric tensors, specifically \begin{align}
\label{Eq:projectiongradient}
\frac{\partial P_{\alpha\beta}^{}}{\partial\mathbf{R}_{j}} ={}& \sum_{I}\frac{\partial P_{\alpha\beta}^{}}{\partial\varphi_{m}^{(I)}\left(\mathbf{r}\right)}\frac{\partial\varphi_{m}^{(I)}\left(\mathbf{r}\right)}{\partial\mathbf{R}_{j}} \\ \nonumber
& +\frac{\partial P_{\alpha\beta}^{}}{\partial O^{mm'}}\frac{\partial O^{mm'}}{\partial\mathbf{R}_{j}},\end{align}
 where, in the tensorial representation~\cite{PhysRevB.83.245124},  
 \begin{align}
\frac{\partial O^{mm'}}{\partial\mathbf{R}_{j}} & {}=-O^{mm''}\frac{\partial O_{m''m''''}^{}}{\partial\mathbf{R}_{j}}O^{m'''m'},\end{align}
 vanishes in the conventional case that  
 the Hubbard projectors are rigidly translated with their host ions.

We note, next, that the Hubbard projectors are usually considered to
be associated with one atomic site only, and so the subspace index $I$ need
only run over subspaces centered on ion $j$. We may  thus suppress
the summation symbol in Eqn.~\ref{Eq:FORCE}, for notational clarity,
since the generalization to multiple subspaces per ion is straightforward.
Denoting the spatial derivative of the Hubbard projectors by the three-component
vector \begin{align}
\mathbf{X}_{\alpha m}^{(j)} & {}=\sum_{I\cap j}\langle\phi_{\alpha}\rvert\mathbf{\nabla}\lvert\varphi_{m}^{(I)}\rangle\\
 & {}=\sum_{I\cap j}\int d\mathbf{r}\;\phi_{\alpha}\left(\mathbf{r}\right)\left[\int d\mathbf{G}\;\left(-i\mathbf{G}\right)e^{-i\mathbf{G}.\mathbf{r}}\varphi_{m}^{(I)}\left(\mathbf{G}\right)\right], \nonumber \end{align}
the remaining terms in Eqn.~\ref{Eq:projectiongradient}
may be expressed as \begin{align}
\frac{\partial P_{\alpha\beta}^{}}{\partial\mathbf{R}_{j}} & {}=\int\left(\frac{\partial}{\partial\varphi_{m}^{(j)}\left(\mathbf{r}\right)}\left[V_{\alpha m'}^{}O^{m'm''}
W_{m''\beta}^{}\right]\right)\\
 & \qquad\times\frac{\partial}{\partial\mathbf{R}}\sum_{I\cap j}\left[\int d\mathbf{G}\; e^{-i\mathbf{G}.\mathbf{r}}\varphi_{m}^{(I)}\left(\mathbf{G}\right)\right]d\mathbf{r}\nonumber \\
 & {}=\int\left(\phi_{\alpha}O^{mm''}W_{m''\beta}^{}+V_{\alpha m'}^{}O^{m'm}\phi_{\beta}\right)\left(\mathbf{r}\right)\nonumber \\
 & \qquad\times\sum_{I\cap j}\left[\int d\mathbf{G}\;\left(-i\mathbf{G}\right)e^{-i\mathbf{G}.\mathbf{r}}\varphi_{m}^{(I)}\left(\mathbf{G}\right)\right]d\mathbf{r}\nonumber \\
 & {}=\mathbf{X}_{\alpha m}^{(j)}O^{mm''}W_{m''\beta}^{}+V_{\alpha m'}^{}O^{m'm}\mathbf{X}_{m\beta}^{(j)\dagger}.
 \nonumber \end{align}

Combining the latter result with Eqs.~\ref{Eq:PROJGRAD}
and~\ref{Eq:FORCE}, 
we conclude
that the tensorially consistent DFT+$U$ contribution to the ionic
forces is succinctly given by the sparse matrix trace,
noting the resemblance to Eqn.~\ref{Eq:gridgradient}, \begin{align}
\mathbf{F}_{U}^{j}=-2
\tilde{K}^{\beta \alpha} V_{\alpha m} H^{m' m} 
\mathbf{X}_{m \beta}^{(j)\dagger}.
\label{Eq:DFTUforce}\end{align}
This may be used, for example, to 
perform DFT+$U$ corrected ionic geometry optimization, 
molecular dynamics, or calculations of vibrational spectra
on large systems.


\begin{thebibliography}{65}%
\makeatletter
\providecommand \@ifxundefined [1]{%
 \@ifx{#1\undefined}
}%
\providecommand \@ifnum [1]{%
 \ifnum #1\expandafter \@firstoftwo
 \else \expandafter \@secondoftwo
 \fi
}%
\providecommand \@ifx [1]{%
 \ifx #1\expandafter \@firstoftwo
 \else \expandafter \@secondoftwo
 \fi
}%
\providecommand \natexlab [1]{#1}%
\providecommand \enquote  [1]{``#1''}%
\providecommand \bibnamefont  [1]{#1}%
\providecommand \bibfnamefont [1]{#1}%
\providecommand \citenamefont [1]{#1}%
\providecommand \href@noop [0]{\@secondoftwo}%
\providecommand \href [0]{\begingroup \@sanitize@url \@href}%
\providecommand \@href[1]{\@@startlink{#1}\@@href}%
\providecommand \@@href[1]{\endgroup#1\@@endlink}%
\providecommand \@sanitize@url [0]{\catcode `\\12\catcode `\$12\catcode
  `\&12\catcode `\#12\catcode `\^12\catcode `\_12\catcode `\%12\relax}%
\providecommand \@@startlink[1]{}%
\providecommand \@@endlink[0]{}%
\providecommand \url  [0]{\begingroup\@sanitize@url \@url }%
\providecommand \@url [1]{\endgroup\@href {#1}{\urlprefix }}%
\providecommand \urlprefix  [0]{URL }%
\providecommand \Eprint [0]{\href }%
\@ifxundefined \urlstyle {%
  \providecommand \doi  [0]{\begingroup \@sanitize@url \@doi}%
  \providecommand \@doi [1]{\endgroup \@@startlink {\doibase
  #1}doi:\discretionary {}{}{}#1\@@endlink }%
}{%
  \providecommand \doi  [0]{doi:\discretionary{}{}{}\begingroup
  \urlstyle{rm}\Url }%
}%
\providecommand \doibase [0]{http://dx.doi.org/}%
\providecommand \Doi [0]{\begingroup \@sanitize@url \@Doi }%
\providecommand \@Doi  [1]{\endgroup\@@startlink{\doibase#1}\@@Doi}%
\providecommand \@@Doi [1]{#1\@@endlink}%
\providecommand \selectlanguage [0]{\@gobble}%
\providecommand \bibinfo  [0]{\@secondoftwo}%
\providecommand \bibfield  [0]{\@secondoftwo}%
\providecommand \translation [1]{[#1]}%
\providecommand \BibitemOpen [0]{}%
\providecommand \bibitemStop [0]{}%
\providecommand \bibitemNoStop [0]{.\EOS\space}%
\providecommand \EOS [0]{\spacefactor3000\relax}%
\providecommand \BibitemShut  [1]{\csname bibitem#1\endcsname}%
\bibitem [{\citenamefont {Hohenberg}\ and\ \citenamefont
  {Kohn}(1964)}]{PhysRev.136.B864}%
  \BibitemOpen
  \bibfield  {author} {\bibinfo {author} {\bibfnamefont {P.}~\bibnamefont
  {Hohenberg}}\ and\ \bibinfo {author} {\bibfnamefont {W.}~\bibnamefont
  {Kohn}},\ }\href@noop {} {\bibfield  {journal} {\bibinfo  {journal} {Phys.
  Rev.},\ }\textbf {\bibinfo {volume} {136}},\ \bibinfo {pages} {B864}
  (\bibinfo {year} {1964})}\BibitemShut {NoStop}%
\bibitem [{\citenamefont {Kohn}\ and\ \citenamefont
  {Sham}(1965)}]{PhysRev.140.A1133}%
  \BibitemOpen
  \bibfield  {author} {\bibinfo {author} {\bibfnamefont {W.}~\bibnamefont
  {Kohn}}\ and\ \bibinfo {author} {\bibfnamefont {L.~J.}\ \bibnamefont
  {Sham}},\ }\href@noop {} {\bibfield  {journal} {\bibinfo  {journal} {Phys.
  Rev.},\ }\textbf {\bibinfo {volume} {140}},\ \bibinfo {pages} {A1133}
  (\bibinfo {year} {1965})}\BibitemShut {NoStop}%
\bibitem [{\citenamefont {Bowler}\ and\ \citenamefont
  {Miyazaki}(2012)}]{bowlerreview}%
  \BibitemOpen
  \bibfield  {author} {\bibinfo {author} {\bibfnamefont {D.~R.}\ \bibnamefont
  {Bowler}}\ and\ \bibinfo {author} {\bibfnamefont {T.}~\bibnamefont
  {Miyazaki}},\ }\href@noop {} {} (\bibinfo {year} {2012}),\ \bibinfo {note}
  {{Rep. Prog. Phys. (in press)}},\ \Eprint
  {http://arxiv.org/abs/arXiv:1108.5976} {arXiv:1108.5976} \BibitemShut
  {NoStop}%
\bibitem [{\citenamefont {Skylaris}\ \emph {et~al.}(2005)\citenamefont
  {Skylaris}, \citenamefont {Haynes}, \citenamefont {Mostofi},\ and\
  \citenamefont {Payne}}]{10.1063/1.1839852}%
  \BibitemOpen
  \bibfield  {author} {\bibinfo {author} {\bibfnamefont {C.-K.}\ \bibnamefont
  {Skylaris}}, \bibinfo {author} {\bibfnamefont {P.~D.}\ \bibnamefont
  {Haynes}}, \bibinfo {author} {\bibfnamefont {A.~A.}\ \bibnamefont {Mostofi}},
  \ and\ \bibinfo {author} {\bibfnamefont {M.~C.}\ \bibnamefont {Payne}},\
  }\href {http://dx.doi.org/10.1063/1.1839852} {\bibfield  {journal} {\bibinfo
  {journal} {J. Chem. Phys.},\ }\textbf {\bibinfo {volume} {122}},\ \bibinfo
  {pages} {084119} (\bibinfo {year} {2005})}\BibitemShut {NoStop}%
\bibitem [{\citenamefont {Hine}\ \emph {et~al.}(2009)\citenamefont {Hine},
  \citenamefont {Haynes}, \citenamefont {Mostofi}, \citenamefont {Skylaris},\
  and\ \citenamefont {Payne}}]{Hine20091041}%
  \BibitemOpen
  \bibfield  {author} {\bibinfo {author} {\bibfnamefont {N.~D.~M.}\
  \bibnamefont {Hine}}, \bibinfo {author} {\bibfnamefont {P.~D.}\ \bibnamefont
  {Haynes}}, \bibinfo {author} {\bibfnamefont {A.~A.}\ \bibnamefont {Mostofi}},
  \bibinfo {author} {\bibfnamefont {C.-K.}\ \bibnamefont {Skylaris}}, \ and\
  \bibinfo {author} {\bibfnamefont {M.~C.}\ \bibnamefont {Payne}},\ }\href@noop
  {} {\bibfield  {journal} {\bibinfo  {journal} {Comp. Phys. Comms.},\ }\textbf
  {\bibinfo {volume} {180}},\ \bibinfo {pages} {1041 } (\bibinfo {year}
  {2009})}\BibitemShut {NoStop}%
\bibitem [{\citenamefont {Hine}\ \emph {et~al.}(2010)\citenamefont {Hine},
  \citenamefont {Haynes}, \citenamefont {Mostofi},\ and\ \citenamefont
  {Payne}}]{hine:114111}%
  \BibitemOpen
  \bibfield  {author} {\bibinfo {author} {\bibfnamefont {N.~D.~M.}\
  \bibnamefont {Hine}}, \bibinfo {author} {\bibfnamefont {P.~D.}\ \bibnamefont
  {Haynes}}, \bibinfo {author} {\bibfnamefont {A.~A.}\ \bibnamefont {Mostofi}},
  \ and\ \bibinfo {author} {\bibfnamefont {M.~C.}\ \bibnamefont {Payne}},\
  }\Doi {10.1063/1.3492379} {\bibfield  {journal} {\bibinfo  {journal} {J.
  Chem. Phys.},\ }\textbf {\bibinfo {volume} {133}},\ \bibinfo {eid} {114111}
  (\bibinfo {year} {2010})}\BibitemShut {NoStop}%
\bibitem [{\citenamefont {Ozaki}\ and\ \citenamefont
  {Kino}(2005)}]{PhysRevB.72.045121}%
  \BibitemOpen
  \bibfield  {author} {\bibinfo {author} {\bibfnamefont {T.}~\bibnamefont
  {Ozaki}}\ and\ \bibinfo {author} {\bibfnamefont {H.}~\bibnamefont {Kino}},\
  }\href {http://link.aps.org/doi/10.1103/PhysRevB.72.045121} {\bibfield
  {journal} {\bibinfo  {journal} {Phys. Rev. B},\ }\textbf {\bibinfo {volume}
  {72}},\ \bibinfo {pages} {045121} (\bibinfo {year} {2005})}\BibitemShut
  {NoStop}%
\bibitem [{\citenamefont {Ozaki}(2010)}]{PhysRevB.82.075131}%
  \BibitemOpen
  \bibfield  {author} {\bibinfo {author} {\bibfnamefont {T.}~\bibnamefont
  {Ozaki}},\ }\Doi {10.1103/PhysRevB.82.075131} {\bibfield  {journal} {\bibinfo
   {journal} {Phys. Rev. B},\ }\textbf {\bibinfo {volume} {82}},\ \bibinfo
  {pages} {075131} (\bibinfo {year} {2010})}\BibitemShut {NoStop}%
\bibitem [{\citenamefont {Bowler}\ \emph {et~al.}(2006)\citenamefont {Bowler},
  \citenamefont {Choudhury}, \citenamefont {Gillan},\ and\ \citenamefont
  {Miyazaki}}]{PSSB:PSSB200541386}%
  \BibitemOpen
  \bibfield  {author} {\bibinfo {author} {\bibfnamefont {D.~R.}\ \bibnamefont
  {Bowler}}, \bibinfo {author} {\bibfnamefont {R.}~\bibnamefont {Choudhury}},
  \bibinfo {author} {\bibfnamefont {M.~J.}\ \bibnamefont {Gillan}}, \ and\
  \bibinfo {author} {\bibfnamefont {T.}~\bibnamefont {Miyazaki}},\ }\href@noop
  {} {\bibfield  {journal} {\bibinfo  {journal} {Phys. Status Solidi B},\
  }\textbf {\bibinfo {volume} {243}},\ \bibinfo {pages} {989} (\bibinfo {year}
  {2006})}\BibitemShut {NoStop}%
\bibitem [{\citenamefont {Bowler}\ and\ \citenamefont
  {Miyazaki}(2010)}]{0953-8984-22-7-074207}%
  \BibitemOpen
  \bibfield  {author} {\bibinfo {author} {\bibfnamefont {D.~R.}\ \bibnamefont
  {Bowler}}\ and\ \bibinfo {author} {\bibfnamefont {T.}~\bibnamefont
  {Miyazaki}},\ }\href {http://stacks.iop.org/0953-8984/22/i=7/a=074207}
  {\bibfield  {journal} {\bibinfo  {journal} {J. Phys.: Condens. Matter},\
  }\textbf {\bibinfo {volume} {22}},\ \bibinfo {pages} {074207} (\bibinfo
  {year} {2010})}\BibitemShut {NoStop}%
\bibitem [{\citenamefont {Anisimov}\ \emph {et~al.}(1991)\citenamefont
  {Anisimov}, \citenamefont {Zaanen},\ and\ \citenamefont
  {Andersen}}]{PhysRevB.44.943}%
  \BibitemOpen
  \bibfield  {author} {\bibinfo {author} {\bibfnamefont {V.~I.}\ \bibnamefont
  {Anisimov}}, \bibinfo {author} {\bibfnamefont {J.}~\bibnamefont {Zaanen}}, \
  and\ \bibinfo {author} {\bibfnamefont {O.~K.}\ \bibnamefont {Andersen}},\
  }\href@noop {} {\bibfield  {journal} {\bibinfo  {journal} {Phys. Rev. B},\
  }\textbf {\bibinfo {volume} {44}},\ \bibinfo {pages} {943} (\bibinfo {year}
  {1991})}\BibitemShut {NoStop}%
\bibitem [{\citenamefont {Anisimov}\ \emph {et~al.}(1993)\citenamefont
  {Anisimov}, \citenamefont {Solovyev}, \citenamefont {Korotin}, \citenamefont
  {Czy\ifmmode~\dot{z}\else \.{z}\fi{}yk},\ and\ \citenamefont
  {Sawatzky}}]{PhysRevB.48.16929}%
  \BibitemOpen
  \bibfield  {author} {\bibinfo {author} {\bibfnamefont {V.~I.}\ \bibnamefont
  {Anisimov}}, \bibinfo {author} {\bibfnamefont {I.~V.}\ \bibnamefont
  {Solovyev}}, \bibinfo {author} {\bibfnamefont {M.~A.}\ \bibnamefont
  {Korotin}}, \bibinfo {author} {\bibfnamefont {M.~T.}\ \bibnamefont
  {Czy\ifmmode~\dot{z}\else \.{z}\fi{}yk}}, \ and\ \bibinfo {author}
  {\bibfnamefont {G.~A.}\ \bibnamefont {Sawatzky}},\ }\href@noop {} {\bibfield
  {journal} {\bibinfo  {journal} {Phys. Rev. B},\ }\textbf {\bibinfo {volume}
  {48}},\ \bibinfo {pages} {16929} (\bibinfo {year} {1993})}\BibitemShut
  {NoStop}%
\bibitem [{\citenamefont {Wierzbowska}\ \emph {et~al.}(2004)\citenamefont
  {Wierzbowska}, \citenamefont {S\'anchez-Portal},\ and\ \citenamefont
  {Sanvito}}]{PhysRevB.70.235209}%
  \BibitemOpen
  \bibfield  {author} {\bibinfo {author} {\bibfnamefont {M.}~\bibnamefont
  {Wierzbowska}}, \bibinfo {author} {\bibfnamefont {D.}~\bibnamefont
  {S\'anchez-Portal}}, \ and\ \bibinfo {author} {\bibfnamefont
  {S.}~\bibnamefont {Sanvito}},\ }\Doi {10.1103/PhysRevB.70.235209} {\bibfield
  {journal} {\bibinfo  {journal} {Phys. Rev. B},\ }\textbf {\bibinfo {volume}
  {70}},\ \bibinfo {pages} {235209} (\bibinfo {year} {2004})}\BibitemShut
  {NoStop}%
\bibitem [{\citenamefont {Han}\ \emph {et~al.}(2006)\citenamefont {Han},
  \citenamefont {Ozaki},\ and\ \citenamefont {Yu}}]{PhysRevB.73.045110}%
  \BibitemOpen
  \bibfield  {author} {\bibinfo {author} {\bibfnamefont {M.~J.}\ \bibnamefont
  {Han}}, \bibinfo {author} {\bibfnamefont {T.}~\bibnamefont {Ozaki}}, \ and\
  \bibinfo {author} {\bibfnamefont {J.}~\bibnamefont {Yu}},\ }\Doi
  {10.1103/PhysRevB.73.045110} {\bibfield  {journal} {\bibinfo  {journal}
  {Phys. Rev. B},\ }\textbf {\bibinfo {volume} {73}},\ \bibinfo {pages}
  {045110} (\bibinfo {year} {2006})}\BibitemShut {NoStop}%
\bibitem [{\citenamefont {Sanna}\ \emph {et~al.}(2007)\citenamefont {Sanna},
  \citenamefont {Hourahine}, \citenamefont {Gerstmann},\ and\ \citenamefont
  {Frauenheim}}]{PhysRevB.76.155128}%
  \BibitemOpen
  \bibfield  {author} {\bibinfo {author} {\bibfnamefont {S.}~\bibnamefont
  {Sanna}}, \bibinfo {author} {\bibfnamefont {B.}~\bibnamefont {Hourahine}},
  \bibinfo {author} {\bibfnamefont {U.}~\bibnamefont {Gerstmann}}, \ and\
  \bibinfo {author} {\bibfnamefont {T.}~\bibnamefont {Frauenheim}},\ }\Doi
  {10.1103/PhysRevB.76.155128} {\bibfield  {journal} {\bibinfo  {journal}
  {Phys. Rev. B},\ }\textbf {\bibinfo {volume} {76}},\ \bibinfo {pages}
  {155128} (\bibinfo {year} {2007})}\BibitemShut {NoStop}%
\bibitem [{\citenamefont {Skylaris}\ \emph {et~al.}(2002)\citenamefont
  {Skylaris}, \citenamefont {Mostofi}, \citenamefont {Haynes}, \citenamefont
  {Di\'eguez},\ and\ \citenamefont {Payne}}]{PhysRevB.66.035119}%
  \BibitemOpen
  \bibfield  {author} {\bibinfo {author} {\bibfnamefont {C.-K.}\ \bibnamefont
  {Skylaris}}, \bibinfo {author} {\bibfnamefont {A.~A.}\ \bibnamefont
  {Mostofi}}, \bibinfo {author} {\bibfnamefont {P.~D.}\ \bibnamefont {Haynes}},
  \bibinfo {author} {\bibfnamefont {O.}~\bibnamefont {Di\'eguez}}, \ and\
  \bibinfo {author} {\bibfnamefont {M.~C.}\ \bibnamefont {Payne}},\ }\Doi
  {10.1103/PhysRevB.66.035119} {\bibfield  {journal} {\bibinfo  {journal}
  {Phys. Rev. B},\ }\textbf {\bibinfo {volume} {66}},\ \bibinfo {pages}
  {035119} (\bibinfo {year} {2002})}\BibitemShut {NoStop}%
\bibitem [{\citenamefont {Perdew}\ and\ \citenamefont
  {Zunger}(1981)}]{PhysRevB.23.5048}%
  \BibitemOpen
  \bibfield  {author} {\bibinfo {author} {\bibfnamefont {J.~P.}\ \bibnamefont
  {Perdew}}\ and\ \bibinfo {author} {\bibfnamefont {A.}~\bibnamefont
  {Zunger}},\ }\href@noop {} {\bibfield  {journal} {\bibinfo  {journal} {Phys.
  Rev. B},\ }\textbf {\bibinfo {volume} {23}},\ \bibinfo {pages} {5048}
  (\bibinfo {year} {1981})}\BibitemShut {NoStop}%
\bibitem [{\citenamefont {Mott}(1949)}]{mott}%
  \BibitemOpen
  \bibfield  {author} {\bibinfo {author} {\bibfnamefont {N.~F.}\ \bibnamefont
  {Mott}},\ }\href@noop {} {\bibfield  {journal} {\bibinfo  {journal} {Proc.
  Phys. Soc.},\ }\textbf {\bibinfo {volume} {A 62}},\ \bibinfo {pages} {416}
  (\bibinfo {year} {1949})}\BibitemShut {NoStop}%
\bibitem [{\citenamefont {Ku}\ \emph {et~al.}(2002)\citenamefont {Ku},
  \citenamefont {Rosner}, \citenamefont {Pickett},\ and\ \citenamefont
  {Scalettar}}]{PhysRevLett.89.167204}%
  \BibitemOpen
  \bibfield  {author} {\bibinfo {author} {\bibfnamefont {W.}~\bibnamefont
  {Ku}}, \bibinfo {author} {\bibfnamefont {H.}~\bibnamefont {Rosner}}, \bibinfo
  {author} {\bibfnamefont {W.~E.}\ \bibnamefont {Pickett}}, \ and\ \bibinfo
  {author} {\bibfnamefont {R.~T.}\ \bibnamefont {Scalettar}},\ }\Doi
  {10.1103/PhysRevLett.89.167204} {\bibfield  {journal} {\bibinfo  {journal}
  {Phys. Rev. Lett.},\ }\textbf {\bibinfo {volume} {89}},\ \bibinfo {pages}
  {167204} (\bibinfo {year} {2002})}\BibitemShut {NoStop}%
\bibitem [{\citenamefont {Mazurenko}\ \emph {et~al.}(2007)\citenamefont
  {Mazurenko}, \citenamefont {Skornyakov}, \citenamefont {Kozhevnikov},
  \citenamefont {Mila},\ and\ \citenamefont {Anisimov}}]{PhysRevB.75.224408}%
  \BibitemOpen
  \bibfield  {author} {\bibinfo {author} {\bibfnamefont {V.~V.}\ \bibnamefont
  {Mazurenko}}, \bibinfo {author} {\bibfnamefont {S.~L.}\ \bibnamefont
  {Skornyakov}}, \bibinfo {author} {\bibfnamefont {A.~V.}\ \bibnamefont
  {Kozhevnikov}}, \bibinfo {author} {\bibfnamefont {F.}~\bibnamefont {Mila}}, \
  and\ \bibinfo {author} {\bibfnamefont {V.~I.}\ \bibnamefont {Anisimov}},\
  }\Doi {10.1103/PhysRevB.75.224408} {\bibfield  {journal} {\bibinfo  {journal}
  {Phys. Rev. B},\ }\textbf {\bibinfo {volume} {75}},\ \bibinfo {pages}
  {224408} (\bibinfo {year} {2007})}\BibitemShut {NoStop}%
\bibitem [{\citenamefont {Yamasaki}\ \emph {et~al.}(2006)\citenamefont
  {Yamasaki}, \citenamefont {Feldbacher}, \citenamefont {Yang}, \citenamefont
  {Andersen},\ and\ \citenamefont {Held}}]{PhysRevLett.96.166401}%
  \BibitemOpen
  \bibfield  {author} {\bibinfo {author} {\bibfnamefont {A.}~\bibnamefont
  {Yamasaki}}, \bibinfo {author} {\bibfnamefont {M.}~\bibnamefont
  {Feldbacher}}, \bibinfo {author} {\bibfnamefont {Y.-F.}\ \bibnamefont
  {Yang}}, \bibinfo {author} {\bibfnamefont {O.~K.}\ \bibnamefont {Andersen}},
  \ and\ \bibinfo {author} {\bibfnamefont {K.}~\bibnamefont {Held}},\ }\Doi
  {10.1103/PhysRevLett.96.166401} {\bibfield  {journal} {\bibinfo  {journal}
  {Phys. Rev. Lett.},\ }\textbf {\bibinfo {volume} {96}},\ \bibinfo {pages}
  {166401} (\bibinfo {year} {2006})}\BibitemShut {NoStop}%
\bibitem [{\citenamefont {Miyake}\ and\ \citenamefont
  {Aryasetiawan}(2008)}]{PhysRevB.77.085122}%
  \BibitemOpen
  \bibfield  {author} {\bibinfo {author} {\bibfnamefont {T.}~\bibnamefont
  {Miyake}}\ and\ \bibinfo {author} {\bibfnamefont {F.}~\bibnamefont
  {Aryasetiawan}},\ }\Doi {10.1103/PhysRevB.77.085122} {\bibfield  {journal}
  {\bibinfo  {journal} {Phys. Rev. B},\ }\textbf {\bibinfo {volume} {77}},\
  \bibinfo {pages} {085122} (\bibinfo {year} {2008})}\BibitemShut {NoStop}%
\bibitem [{\citenamefont {O'Regan}\ \emph {et~al.}(2010)\citenamefont
  {O'Regan}, \citenamefont {Hine}, \citenamefont {Payne},\ and\ \citenamefont
  {Mostofi}}]{PhysRevB.82.081102}%
  \BibitemOpen
  \bibfield  {author} {\bibinfo {author} {\bibfnamefont {D.~D.}\ \bibnamefont
  {O'Regan}}, \bibinfo {author} {\bibfnamefont {N.~D.~M.}\ \bibnamefont
  {Hine}}, \bibinfo {author} {\bibfnamefont {M.~C.}\ \bibnamefont {Payne}}, \
  and\ \bibinfo {author} {\bibfnamefont {A.~A.}\ \bibnamefont {Mostofi}},\
  }\Doi {10.1103/PhysRevB.82.081102} {\bibfield  {journal} {\bibinfo  {journal}
  {Phys. Rev. B},\ }\textbf {\bibinfo {volume} {82}},\ \bibinfo {pages}
  {081102} (\bibinfo {year} {2010})}\BibitemShut {NoStop}%
\bibitem [{\citenamefont {O'Regan}\ \emph {et~al.}(2011)\citenamefont
  {O'Regan}, \citenamefont {Payne},\ and\ \citenamefont
  {Mostofi}}]{PhysRevB.83.245124}%
  \BibitemOpen
  \bibfield  {author} {\bibinfo {author} {\bibfnamefont {D.~D.}\ \bibnamefont
  {O'Regan}}, \bibinfo {author} {\bibfnamefont {M.~C.}\ \bibnamefont {Payne}},
  \ and\ \bibinfo {author} {\bibfnamefont {A.~A.}\ \bibnamefont {Mostofi}},\
  }\Doi {10.1103/PhysRevB.83.245124} {\bibfield  {journal} {\bibinfo  {journal}
  {Phys. Rev. B},\ }\textbf {\bibinfo {volume} {83}},\ \bibinfo {pages}
  {245124} (\bibinfo {year} {2011})}\BibitemShut {NoStop}%
\bibitem [{Note1()}]{Note1}%
  \BibitemOpen
  \bibinfo {note} {We employ the Einstein convention, where pairs of identical
  indices are summed over unless in parentheses.}\BibitemShut {Stop}%
\bibitem [{\citenamefont {Artacho}\ and\ \citenamefont {Mil\'{a}ns~del
  Bosch}(1991)}]{Artacho}%
  \BibitemOpen
  \bibfield  {author} {\bibinfo {author} {\bibfnamefont {E.}~\bibnamefont
  {Artacho}}\ and\ \bibinfo {author} {\bibfnamefont {L.}~\bibnamefont
  {Mil\'{a}ns~del Bosch}},\ }\href@noop {} {\bibfield  {journal} {\bibinfo
  {journal} {Phys. Rev. A},\ }\textbf {\bibinfo {volume} {43}},\ \bibinfo
  {pages} {5770} (\bibinfo {year} {1991})}\BibitemShut {NoStop}%
\bibitem [{\citenamefont {Anisimov}\ \emph {et~al.}(1997)\citenamefont
  {Anisimov}, \citenamefont {Aryasetiawan},\ and\ \citenamefont
  {Lichtenstein}}]{0953-8984-9-4-002}%
  \BibitemOpen
  \bibfield  {author} {\bibinfo {author} {\bibfnamefont {V.~I.}\ \bibnamefont
  {Anisimov}}, \bibinfo {author} {\bibfnamefont {F.}~\bibnamefont
  {Aryasetiawan}}, \ and\ \bibinfo {author} {\bibfnamefont {A.~I.}\
  \bibnamefont {Lichtenstein}},\ }\href
  {http://stacks.iop.org/0953-8984/9/i=4/a=002} {\bibfield  {journal} {\bibinfo
   {journal} {J. Phys.: Condens. Matter},\ }\textbf {\bibinfo {volume} {9}},\
  \bibinfo {pages} {767} (\bibinfo {year} {1997})}\BibitemShut {NoStop}%
\bibitem [{\citenamefont {Liechtenstein}\ \emph {et~al.}(1995)\citenamefont
  {Liechtenstein}, \citenamefont {Anisimov},\ and\ \citenamefont
  {Zaanen}}]{PhysRevB.52.R5467}%
  \BibitemOpen
  \bibfield  {author} {\bibinfo {author} {\bibfnamefont {A.~I.}\ \bibnamefont
  {Liechtenstein}}, \bibinfo {author} {\bibfnamefont {V.~I.}\ \bibnamefont
  {Anisimov}}, \ and\ \bibinfo {author} {\bibfnamefont {J.}~\bibnamefont
  {Zaanen}},\ }\href@noop {} {\bibfield  {journal} {\bibinfo  {journal} {Phys.
  Rev. B},\ }\textbf {\bibinfo {volume} {52}},\ \bibinfo {pages} {R5467}
  (\bibinfo {year} {1995})}\BibitemShut {NoStop}%
\bibitem [{\citenamefont {Cococcioni}\ and\ \citenamefont
  {de~Gironcoli}(2005)}]{PhysRevB.71.035105}%
  \BibitemOpen
  \bibfield  {author} {\bibinfo {author} {\bibfnamefont {M.}~\bibnamefont
  {Cococcioni}}\ and\ \bibinfo {author} {\bibfnamefont {S.}~\bibnamefont
  {de~Gironcoli}},\ }\href@noop {} {\bibfield  {journal} {\bibinfo  {journal}
  {Phys. Rev. B},\ }\textbf {\bibinfo {volume} {71}},\ \bibinfo {pages}
  {035105} (\bibinfo {year} {2005})}\BibitemShut {NoStop}%
\bibitem [{\citenamefont {Brouder}\ \emph {et~al.}(2007)\citenamefont
  {Brouder}, \citenamefont {Panati}, \citenamefont {Calandra}, \citenamefont
  {Mourougane},\ and\ \citenamefont {Marzari}}]{PhysRevLett.98.046402}%
  \BibitemOpen
  \bibfield  {author} {\bibinfo {author} {\bibfnamefont {C.}~\bibnamefont
  {Brouder}}, \bibinfo {author} {\bibfnamefont {G.}~\bibnamefont {Panati}},
  \bibinfo {author} {\bibfnamefont {M.}~\bibnamefont {Calandra}}, \bibinfo
  {author} {\bibfnamefont {C.}~\bibnamefont {Mourougane}}, \ and\ \bibinfo
  {author} {\bibfnamefont {N.}~\bibnamefont {Marzari}},\ }\Doi
  {10.1103/PhysRevLett.98.046402} {\bibfield  {journal} {\bibinfo  {journal}
  {Phys. Rev. Lett.},\ }\textbf {\bibinfo {volume} {98}},\ \bibinfo {pages}
  {046402} (\bibinfo {year} {2007})}\BibitemShut {NoStop}%
\bibitem [{\citenamefont {Campo~Jr.}\ and\ \citenamefont
  {Cococcioni}(2010)}]{0953-8984-22-5-055602}%
  \BibitemOpen
  \bibfield  {author} {\bibinfo {author} {\bibfnamefont {V.~L.}\ \bibnamefont
  {Campo~Jr.}}\ and\ \bibinfo {author} {\bibfnamefont {M.}~\bibnamefont
  {Cococcioni}},\ }\href {http://stacks.iop.org/0953-8984/22/i=5/a=055602}
  {\bibfield  {journal} {\bibinfo  {journal} {J. Phys.: Condens. Matter},\
  }\textbf {\bibinfo {volume} {22}},\ \bibinfo {pages} {055602} (\bibinfo
  {year} {2010})}\BibitemShut {NoStop}%
\bibitem [{\citenamefont {Ozaki}(2003)}]{PhysRevB.67.155108}%
  \BibitemOpen
  \bibfield  {author} {\bibinfo {author} {\bibfnamefont {T.}~\bibnamefont
  {Ozaki}},\ }\Doi {10.1103/PhysRevB.67.155108} {\bibfield  {journal} {\bibinfo
   {journal} {Phys. Rev. B},\ }\textbf {\bibinfo {volume} {67}},\ \bibinfo
  {pages} {155108} (\bibinfo {year} {2003})}\BibitemShut {NoStop}%
\bibitem [{\citenamefont {Ozaki}\ and\ \citenamefont
  {Kino}(2004){\natexlab{a}}}]{PhysRevB.69.195113}%
  \BibitemOpen
  \bibfield  {author} {\bibinfo {author} {\bibfnamefont {T.}~\bibnamefont
  {Ozaki}}\ and\ \bibinfo {author} {\bibfnamefont {H.}~\bibnamefont {Kino}},\
  }\Doi {10.1103/PhysRevB.69.195113} {\bibfield  {journal} {\bibinfo  {journal}
  {Phys. Rev. B},\ }\textbf {\bibinfo {volume} {69}},\ \bibinfo {pages}
  {195113} (\bibinfo {year} {2004}{\natexlab{a}})}\BibitemShut {NoStop}%
\bibitem [{\citenamefont {Ozaki}\ and\ \citenamefont
  {Kino}(2004){\natexlab{b}}}]{ozaki:10879}%
  \BibitemOpen
  \bibfield  {author} {\bibinfo {author} {\bibfnamefont {T.}~\bibnamefont
  {Ozaki}}\ and\ \bibinfo {author} {\bibfnamefont {H.}~\bibnamefont {Kino}},\
  }\Doi {10.1063/1.1794591} {\bibfield  {journal} {\bibinfo  {journal} {J.
  Chem. Phys.},\ }\textbf {\bibinfo {volume} {121}},\ \bibinfo {pages} {10879}
  (\bibinfo {year} {2004}{\natexlab{b}})}\BibitemShut {NoStop}%
\bibitem [{\citenamefont {Mauri}\ \emph {et~al.}(1993)\citenamefont {Mauri},
  \citenamefont {Galli},\ and\ \citenamefont {Car}}]{PhysRevB.47.9973}%
  \BibitemOpen
  \bibfield  {author} {\bibinfo {author} {\bibfnamefont {F.}~\bibnamefont
  {Mauri}}, \bibinfo {author} {\bibfnamefont {G.}~\bibnamefont {Galli}}, \ and\
  \bibinfo {author} {\bibfnamefont {R.}~\bibnamefont {Car}},\ }\href@noop {}
  {\bibfield  {journal} {\bibinfo  {journal} {Phys. Rev. B},\ }\textbf
  {\bibinfo {volume} {47}},\ \bibinfo {pages} {9973} (\bibinfo {year}
  {1993})}\BibitemShut {NoStop}%
\bibitem [{\citenamefont {Mauri}\ and\ \citenamefont
  {Galli}(1994)}]{PhysRevB.50.4316}%
  \BibitemOpen
  \bibfield  {author} {\bibinfo {author} {\bibfnamefont {F.}~\bibnamefont
  {Mauri}}\ and\ \bibinfo {author} {\bibfnamefont {G.}~\bibnamefont {Galli}},\
  }\href@noop {} {\bibfield  {journal} {\bibinfo  {journal} {Phys. Rev. B},\
  }\textbf {\bibinfo {volume} {50}},\ \bibinfo {pages} {4316} (\bibinfo {year}
  {1994})}\BibitemShut {NoStop}%
\bibitem [{\citenamefont {Kim}\ \emph {et~al.}(1995)\citenamefont {Kim},
  \citenamefont {Mauri},\ and\ \citenamefont {Galli}}]{PhysRevB.52.1640}%
  \BibitemOpen
  \bibfield  {author} {\bibinfo {author} {\bibfnamefont {J.}~\bibnamefont
  {Kim}}, \bibinfo {author} {\bibfnamefont {F.}~\bibnamefont {Mauri}}, \ and\
  \bibinfo {author} {\bibfnamefont {G.}~\bibnamefont {Galli}},\ }\href@noop {}
  {\bibfield  {journal} {\bibinfo  {journal} {Phys. Rev. B},\ }\textbf
  {\bibinfo {volume} {52}},\ \bibinfo {pages} {1640} (\bibinfo {year}
  {1995})}\BibitemShut {NoStop}%
\bibitem [{\citenamefont {Ordej\'on}\ \emph {et~al.}(1995)\citenamefont
  {Ordej\'on}, \citenamefont {Drabold}, \citenamefont {Martin},\ and\
  \citenamefont {Grumbach}}]{PhysRevB.51.1456}%
  \BibitemOpen
  \bibfield  {author} {\bibinfo {author} {\bibfnamefont {P.}~\bibnamefont
  {Ordej\'on}}, \bibinfo {author} {\bibfnamefont {D.~A.}\ \bibnamefont
  {Drabold}}, \bibinfo {author} {\bibfnamefont {R.~M.}\ \bibnamefont {Martin}},
  \ and\ \bibinfo {author} {\bibfnamefont {M.~P.}\ \bibnamefont {Grumbach}},\
  }\href@noop {} {\bibfield  {journal} {\bibinfo  {journal} {Phys. Rev. B},\
  }\textbf {\bibinfo {volume} {51}},\ \bibinfo {pages} {1456} (\bibinfo {year}
  {1995})}\BibitemShut {NoStop}%
\bibitem [{\citenamefont {Ratcliff}\ \emph {et~al.}(2011)\citenamefont
  {Ratcliff}, \citenamefont {Hine},\ and\ \citenamefont
  {Haynes}}]{PhysRevB.84.165131}%
  \BibitemOpen
  \bibfield  {author} {\bibinfo {author} {\bibfnamefont {L.~E.}\ \bibnamefont
  {Ratcliff}}, \bibinfo {author} {\bibfnamefont {N.~D.~M.}\ \bibnamefont
  {Hine}}, \ and\ \bibinfo {author} {\bibfnamefont {P.~D.}\ \bibnamefont
  {Haynes}},\ }\Doi {10.1103/PhysRevB.84.165131} {\bibfield  {journal}
  {\bibinfo  {journal} {Phys. Rev. B},\ }\textbf {\bibinfo {volume} {84}},\
  \bibinfo {pages} {165131} (\bibinfo {year} {2011})}\BibitemShut {NoStop}%
\bibitem [{\citenamefont {Scherlis}\ \emph {et~al.}(2007)\citenamefont
  {Scherlis}, \citenamefont {Cococcioni}, \citenamefont {Sit},\ and\
  \citenamefont {Marzari}}]{doi:10.1021/jp070549l}%
  \BibitemOpen
  \bibfield  {author} {\bibinfo {author} {\bibfnamefont {D.~A.}\ \bibnamefont
  {Scherlis}}, \bibinfo {author} {\bibfnamefont {M.}~\bibnamefont
  {Cococcioni}}, \bibinfo {author} {\bibfnamefont {P.}~\bibnamefont {Sit}}, \
  and\ \bibinfo {author} {\bibfnamefont {N.}~\bibnamefont {Marzari}},\
  }\href@noop {} {\bibfield  {journal} {\bibinfo  {journal} {J. Phys. Chem.
  B},\ }\textbf {\bibinfo {volume} {111}},\ \bibinfo {pages} {7384} (\bibinfo
  {year} {2007})}\BibitemShut {NoStop}%
\bibitem [{\citenamefont {Kulik}\ and\ \citenamefont
  {Marzari}(2008)}]{kulik:134314}%
  \BibitemOpen
  \bibfield  {author} {\bibinfo {author} {\bibfnamefont {H.~J.}\ \bibnamefont
  {Kulik}}\ and\ \bibinfo {author} {\bibfnamefont {N.}~\bibnamefont
  {Marzari}},\ }\href {http://link.aip.org/link/?JCP/129/134314/1} {\bibfield
  {journal} {\bibinfo  {journal} {J. Chem. Phys.},\ }\textbf {\bibinfo {volume}
  {129}},\ \bibinfo {eid} {134314} (\bibinfo {year} {2008})}\BibitemShut
  {NoStop}%
\bibitem [{\citenamefont {Kulik}\ and\ \citenamefont
  {Marzari}(2010)}]{kulik:114103}%
  \BibitemOpen
  \bibfield  {author} {\bibinfo {author} {\bibfnamefont {H.~J.}\ \bibnamefont
  {Kulik}}\ and\ \bibinfo {author} {\bibfnamefont {N.}~\bibnamefont
  {Marzari}},\ }\href {http://link.aip.org/link/?JCP/133/114103/1} {\bibfield
  {journal} {\bibinfo  {journal} {J. Chem. Phys},\ }\textbf {\bibinfo {volume}
  {133}},\ \bibinfo {eid} {114103} (\bibinfo {year} {2010})}\BibitemShut
  {NoStop}%
\bibitem [{\citenamefont {Kulik}\ and\ \citenamefont
  {Marzari}(2011)}]{kulik:094103}%
  \BibitemOpen
  \bibfield  {author} {\bibinfo {author} {\bibfnamefont {H.~J.}\ \bibnamefont
  {Kulik}}\ and\ \bibinfo {author} {\bibfnamefont {N.}~\bibnamefont
  {Marzari}},\ }\Doi {10.1063/1.3559452} {\bibfield  {journal} {\bibinfo
  {journal} {J. Chem. Phys.},\ }\textbf {\bibinfo {volume} {134}},\ \bibinfo
  {eid} {094103} (\bibinfo {year} {2011})}\BibitemShut {NoStop}%
\bibitem [{\citenamefont {Hsu}\ \emph {et~al.}(2009)\citenamefont {Hsu},
  \citenamefont {Umemoto}, \citenamefont {Cococcioni},\ and\ \citenamefont
  {Wentzcovitch}}]{PhysRevB.79.125124}%
  \BibitemOpen
  \bibfield  {author} {\bibinfo {author} {\bibfnamefont {H.}~\bibnamefont
  {Hsu}}, \bibinfo {author} {\bibfnamefont {K.}~\bibnamefont {Umemoto}},
  \bibinfo {author} {\bibfnamefont {M.}~\bibnamefont {Cococcioni}}, \ and\
  \bibinfo {author} {\bibfnamefont {R.}~\bibnamefont {Wentzcovitch}},\
  }\href@noop {} {\bibfield  {journal} {\bibinfo  {journal} {Phys. Rev. B},\
  }\textbf {\bibinfo {volume} {79}},\ \bibinfo {pages} {125124} (\bibinfo
  {year} {2009})}\BibitemShut {NoStop}%
\bibitem [{\citenamefont {Himmetoglu}\ \emph {et~al.}(2011)\citenamefont
  {Himmetoglu}, \citenamefont {Wentzcovitch},\ and\ \citenamefont
  {Cococcioni}}]{PhysRevB.84.115108}%
  \BibitemOpen
  \bibfield  {author} {\bibinfo {author} {\bibfnamefont {B.}~\bibnamefont
  {Himmetoglu}}, \bibinfo {author} {\bibfnamefont {R.~M.}\ \bibnamefont
  {Wentzcovitch}}, \ and\ \bibinfo {author} {\bibfnamefont {M.}~\bibnamefont
  {Cococcioni}},\ }\Doi {10.1103/PhysRevB.84.115108} {\bibfield  {journal}
  {\bibinfo  {journal} {Phys. Rev. B},\ }\textbf {\bibinfo {volume} {84}},\
  \bibinfo {pages} {115108} (\bibinfo {year} {2011})}\BibitemShut {NoStop}%
\bibitem [{\citenamefont {Hine}\ \emph {et~al.}(2011)\citenamefont {Hine},
  \citenamefont {Robinson}, \citenamefont {Haynes}, \citenamefont {Skylaris},
  \citenamefont {Payne},\ and\ \citenamefont {Mostofi}}]{PhysRevB.83.195102}%
  \BibitemOpen
  \bibfield  {author} {\bibinfo {author} {\bibfnamefont {N.~D.~M.}\
  \bibnamefont {Hine}}, \bibinfo {author} {\bibfnamefont {M.}~\bibnamefont
  {Robinson}}, \bibinfo {author} {\bibfnamefont {P.~D.}\ \bibnamefont
  {Haynes}}, \bibinfo {author} {\bibfnamefont {C.-K.}\ \bibnamefont
  {Skylaris}}, \bibinfo {author} {\bibfnamefont {M.~C.}\ \bibnamefont {Payne}},
  \ and\ \bibinfo {author} {\bibfnamefont {A.~A.}\ \bibnamefont {Mostofi}},\
  }\Doi {10.1103/PhysRevB.83.195102} {\bibfield  {journal} {\bibinfo  {journal}
  {Phys. Rev. B},\ }\textbf {\bibinfo {volume} {83}},\ \bibinfo {pages}
  {195102} (\bibinfo {year} {2011})}\BibitemShut {NoStop}%
\bibitem [{\citenamefont {Li}\ \emph {et~al.}(1993)\citenamefont {Li},
  \citenamefont {Nunes},\ and\ \citenamefont {Vanderbilt}}]{PhysRevB.47.10891}%
  \BibitemOpen
  \bibfield  {author} {\bibinfo {author} {\bibfnamefont {X.-P.}\ \bibnamefont
  {Li}}, \bibinfo {author} {\bibfnamefont {R.~W.}\ \bibnamefont {Nunes}}, \
  and\ \bibinfo {author} {\bibfnamefont {D.}~\bibnamefont {Vanderbilt}},\
  }\href@noop {} {\bibfield  {journal} {\bibinfo  {journal} {Phys. Rev. B},\
  }\textbf {\bibinfo {volume} {47}},\ \bibinfo {pages} {10891} (\bibinfo {year}
  {1993})}\BibitemShut {NoStop}%
\bibitem [{\citenamefont {Nunes}\ and\ \citenamefont
  {Vanderbilt}(1994)}]{PhysRevB.50.17611}%
  \BibitemOpen
  \bibfield  {author} {\bibinfo {author} {\bibfnamefont {R.~W.}\ \bibnamefont
  {Nunes}}\ and\ \bibinfo {author} {\bibfnamefont {D.}~\bibnamefont
  {Vanderbilt}},\ }\href@noop {} {\bibfield  {journal} {\bibinfo  {journal}
  {Phys. Rev. B},\ }\textbf {\bibinfo {volume} {50}},\ \bibinfo {pages} {17611}
  (\bibinfo {year} {1994})}\BibitemShut {NoStop}%
\bibitem [{\citenamefont {Daw}(1993)}]{PhysRevB.47.10895}%
  \BibitemOpen
  \bibfield  {author} {\bibinfo {author} {\bibfnamefont {M.~S.}\ \bibnamefont
  {Daw}},\ }\href@noop {} {\bibfield  {journal} {\bibinfo  {journal} {Phys.
  Rev. B},\ }\textbf {\bibinfo {volume} {47}},\ \bibinfo {pages} {10895}
  (\bibinfo {year} {1993})}\BibitemShut {NoStop}%
\bibitem [{\citenamefont {White}\ \emph {et~al.}(1997)\citenamefont {White},
  \citenamefont {Maslen}, \citenamefont {Lee},\ and\ \citenamefont
  {Head-Gordon}}]{White1997133}%
  \BibitemOpen
  \bibfield  {author} {\bibinfo {author} {\bibfnamefont {C.~A.}\ \bibnamefont
  {White}}, \bibinfo {author} {\bibfnamefont {P.}~\bibnamefont {Maslen}},
  \bibinfo {author} {\bibfnamefont {M.~S.}\ \bibnamefont {Lee}}, \ and\
  \bibinfo {author} {\bibfnamefont {M.}~\bibnamefont {Head-Gordon}},\
  }\href@noop {} {\bibfield  {journal} {\bibinfo  {journal} {Chem. Phys.
  Lett.},\ }\textbf {\bibinfo {volume} {276}},\ \bibinfo {pages} {133 }
  (\bibinfo {year} {1997})}\BibitemShut {NoStop}%
\bibitem [{\citenamefont {Towler}\ \emph {et~al.}(1994)\citenamefont {Towler},
  \citenamefont {Allan}, \citenamefont {Harrison}, \citenamefont {Saunders},
  \citenamefont {Mackrodt},\ and\ \citenamefont {Apr\`a}}]{PhysRevB.50.5041}%
  \BibitemOpen
  \bibfield  {author} {\bibinfo {author} {\bibfnamefont {M.~D.}\ \bibnamefont
  {Towler}}, \bibinfo {author} {\bibfnamefont {N.~L.}\ \bibnamefont {Allan}},
  \bibinfo {author} {\bibfnamefont {N.~M.}\ \bibnamefont {Harrison}}, \bibinfo
  {author} {\bibfnamefont {V.~R.}\ \bibnamefont {Saunders}}, \bibinfo {author}
  {\bibfnamefont {W.~C.}\ \bibnamefont {Mackrodt}}, \ and\ \bibinfo {author}
  {\bibfnamefont {E.}~\bibnamefont {Apr\`a}},\ }\href@noop {} {\bibfield
  {journal} {\bibinfo  {journal} {Phys. Rev. B},\ }\textbf {\bibinfo {volume}
  {50}},\ \bibinfo {pages} {5041} (\bibinfo {year} {1994})}\BibitemShut
  {NoStop}%
\bibitem [{\citenamefont {Sawatzky}\ and\ \citenamefont
  {Allen}(1984)}]{PhysRevLett.53.2339}%
  \BibitemOpen
  \bibfield  {author} {\bibinfo {author} {\bibfnamefont {G.~A.}\ \bibnamefont
  {Sawatzky}}\ and\ \bibinfo {author} {\bibfnamefont {J.~W.}\ \bibnamefont
  {Allen}},\ }\href@noop {} {\bibfield  {journal} {\bibinfo  {journal} {Phys.
  Rev. Lett.},\ }\textbf {\bibinfo {volume} {53}},\ \bibinfo {pages} {2339}
  (\bibinfo {year} {1984})}\BibitemShut {NoStop}%
\bibitem [{\citenamefont {Ren}\ \emph {et~al.}(2006)\citenamefont {Ren},
  \citenamefont {Leonov}, \citenamefont {Keller}, \citenamefont {Kollar},
  \citenamefont {Nekrasov},\ and\ \citenamefont
  {Vollhardt}}]{PhysRevB.74.195114}%
  \BibitemOpen
  \bibfield  {author} {\bibinfo {author} {\bibfnamefont {X.}~\bibnamefont
  {Ren}}, \bibinfo {author} {\bibfnamefont {I.}~\bibnamefont {Leonov}},
  \bibinfo {author} {\bibfnamefont {G.}~\bibnamefont {Keller}}, \bibinfo
  {author} {\bibfnamefont {M.}~\bibnamefont {Kollar}}, \bibinfo {author}
  {\bibfnamefont {I.}~\bibnamefont {Nekrasov}}, \ and\ \bibinfo {author}
  {\bibfnamefont {D.}~\bibnamefont {Vollhardt}},\ }\href@noop {} {\bibfield
  {journal} {\bibinfo  {journal} {Phys. Rev. B},\ }\textbf {\bibinfo {volume}
  {74}},\ \bibinfo {pages} {195114} (\bibinfo {year} {2006})}\BibitemShut
  {NoStop}%
\bibitem [{\citenamefont {Svane}\ and\ \citenamefont
  {Gunnarsson}(1990)}]{PhysRevLett.65.1148}%
  \BibitemOpen
  \bibfield  {author} {\bibinfo {author} {\bibfnamefont {A.}~\bibnamefont
  {Svane}}\ and\ \bibinfo {author} {\bibfnamefont {O.}~\bibnamefont
  {Gunnarsson}},\ }\href@noop {} {\bibfield  {journal} {\bibinfo  {journal}
  {Phys. Rev. Lett.},\ }\textbf {\bibinfo {volume} {65}},\ \bibinfo {pages}
  {1148} (\bibinfo {year} {1990})}\BibitemShut {NoStop}%
\bibitem [{\citenamefont {Dudarev}\ \emph {et~al.}(1998)\citenamefont
  {Dudarev}, \citenamefont {Botton}, \citenamefont {Savrasov}, \citenamefont
  {Humphreys},\ and\ \citenamefont {Sutton}}]{PhysRevB.57.1505}%
  \BibitemOpen
  \bibfield  {author} {\bibinfo {author} {\bibfnamefont {S.~L.}\ \bibnamefont
  {Dudarev}}, \bibinfo {author} {\bibfnamefont {G.~A.}\ \bibnamefont {Botton}},
  \bibinfo {author} {\bibfnamefont {S.~Y.}\ \bibnamefont {Savrasov}}, \bibinfo
  {author} {\bibfnamefont {C.~J.}\ \bibnamefont {Humphreys}}, \ and\ \bibinfo
  {author} {\bibfnamefont {A.~P.}\ \bibnamefont {Sutton}},\ }\href@noop {}
  {\bibfield  {journal} {\bibinfo  {journal} {Phys. Rev. B},\ }\textbf
  {\bibinfo {volume} {57}},\ \bibinfo {pages} {1505} (\bibinfo {year}
  {1998})}\BibitemShut {NoStop}%
\bibitem [{\citenamefont {Bengone}\ \emph {et~al.}(2000)\citenamefont
  {Bengone}, \citenamefont {Alouani}, \citenamefont {Bl\"ochl},\ and\
  \citenamefont {Hugel}}]{PhysRevB.62.16392}%
  \BibitemOpen
  \bibfield  {author} {\bibinfo {author} {\bibfnamefont {O.}~\bibnamefont
  {Bengone}}, \bibinfo {author} {\bibfnamefont {M.}~\bibnamefont {Alouani}},
  \bibinfo {author} {\bibfnamefont {P.}~\bibnamefont {Bl\"ochl}}, \ and\
  \bibinfo {author} {\bibfnamefont {J.}~\bibnamefont {Hugel}},\ }\href@noop {}
  {\bibfield  {journal} {\bibinfo  {journal} {Phys. Rev. B},\ }\textbf
  {\bibinfo {volume} {62}},\ \bibinfo {pages} {16392} (\bibinfo {year}
  {2000})}\BibitemShut {NoStop}%
\bibitem [{\citenamefont {Pickett}\ \emph {et~al.}(1998)\citenamefont
  {Pickett}, \citenamefont {Erwin},\ and\ \citenamefont
  {Ethridge}}]{PhysRevB.58.1201}%
  \BibitemOpen
  \bibfield  {author} {\bibinfo {author} {\bibfnamefont {W.~E.}\ \bibnamefont
  {Pickett}}, \bibinfo {author} {\bibfnamefont {S.~C.}\ \bibnamefont {Erwin}},
  \ and\ \bibinfo {author} {\bibfnamefont {E.~C.}\ \bibnamefont {Ethridge}},\
  }\href@noop {} {\bibfield  {journal} {\bibinfo  {journal} {Phys. Rev. B},\
  }\textbf {\bibinfo {volume} {58}},\ \bibinfo {pages} {1201} (\bibinfo {year}
  {1998})}\BibitemShut {NoStop}%
\bibitem [{\citenamefont {L\'opez}\ \emph {et~al.}(2009)\citenamefont
  {L\'opez}, \citenamefont {Romero}, \citenamefont {Mej\'\i{}a-L\'opez},
  \citenamefont {Mazo-Zuluaga},\ and\ \citenamefont
  {Restrepo}}]{PhysRevB.80.085107}%
  \BibitemOpen
  \bibfield  {author} {\bibinfo {author} {\bibfnamefont {S.}~\bibnamefont
  {L\'opez}}, \bibinfo {author} {\bibfnamefont {A.~H.}\ \bibnamefont {Romero}},
  \bibinfo {author} {\bibfnamefont {J.}~\bibnamefont {Mej\'\i{}a-L\'opez}},
  \bibinfo {author} {\bibfnamefont {J.}~\bibnamefont {Mazo-Zuluaga}}, \ and\
  \bibinfo {author} {\bibfnamefont {J.}~\bibnamefont {Restrepo}},\ }\href@noop
  {} {\bibfield  {journal} {\bibinfo  {journal} {Phys. Rev. B},\ }\textbf
  {\bibinfo {volume} {80}},\ \bibinfo {pages} {085107} (\bibinfo {year}
  {2009})}\BibitemShut {NoStop}%
\bibitem [{\citenamefont {Palot\'as}\ \emph {et~al.}(2010)\citenamefont
  {Palot\'as}, \citenamefont {Andriotis},\ and\ \citenamefont
  {Lappas}}]{PhysRevB.81.075403}%
  \BibitemOpen
  \bibfield  {author} {\bibinfo {author} {\bibfnamefont {K.}~\bibnamefont
  {Palot\'as}}, \bibinfo {author} {\bibfnamefont {A.~N.}\ \bibnamefont
  {Andriotis}}, \ and\ \bibinfo {author} {\bibfnamefont {A.}~\bibnamefont
  {Lappas}},\ }\href@noop {} {\bibfield  {journal} {\bibinfo  {journal} {Phys.
  Rev. B},\ }\textbf {\bibinfo {volume} {81}},\ \bibinfo {pages} {075403}
  (\bibinfo {year} {2010})}\BibitemShut {NoStop}%
\bibitem [{opi()}]{opium}%
  \BibitemOpen
  \href@noop {} {}\bibinfo {note} {Norm-conserving pseudopotentials,
  relativistically corrected and with a non-linear core correction for nickel,
  were generated using the Opium code available at {\tt
  http://opium.sourceforge.net}.}\BibitemShut {Stop}%
\bibitem [{\citenamefont {Ozaki}(2001)}]{PhysRevB.64.195110}%
  \BibitemOpen
  \bibfield  {author} {\bibinfo {author} {\bibfnamefont {T.}~\bibnamefont
  {Ozaki}},\ }\href@noop {} {\bibfield  {journal} {\bibinfo  {journal} {Phys.
  Rev. B},\ }\textbf {\bibinfo {volume} {64}},\ \bibinfo {pages} {195110}
  (\bibinfo {year} {2001})}\BibitemShut {NoStop}%
\bibitem [{\citenamefont {O'Regan}\ \emph {et~al.}()\citenamefont {O'Regan},
  \citenamefont {Payne},\ and\ \citenamefont {Mostofi}}]{forthcomingGaMnAs}%
  \BibitemOpen
  \bibfield  {author} {\bibinfo {author} {\bibfnamefont {D.~D.}\ \bibnamefont
  {O'Regan}}, \bibinfo {author} {\bibfnamefont {M.~C.}\ \bibnamefont {Payne}},
  \ and\ \bibinfo {author} {\bibfnamefont {A.~A.}\ \bibnamefont {Mostofi}},\
  }\href@noop {} {}\bibinfo {note} {In preparation (2012)}\BibitemShut
  {NoStop}%
\bibitem [{\citenamefont {Weber}\ \emph {et~al.}()\citenamefont {Weber},
  \citenamefont {O'Regan}, \citenamefont {Hine}, \citenamefont {Payne},
  \citenamefont {Kotliar},\ and\ \citenamefont {Littlewood}}]{forthcomingVO2}%
  \BibitemOpen
  \bibfield  {author} {\bibinfo {author} {\bibfnamefont {C.}~\bibnamefont
  {Weber}}, \bibinfo {author} {\bibfnamefont {D.~D.}\ \bibnamefont {O'Regan}},
  \bibinfo {author} {\bibfnamefont {N.~D.~M.}\ \bibnamefont {Hine}}, \bibinfo
  {author} {\bibfnamefont {M.~C.}\ \bibnamefont {Payne}}, \bibinfo {author}
  {\bibfnamefont {G.}~\bibnamefont {Kotliar}}, \ and\ \bibinfo {author}
  {\bibfnamefont {P.~B.}\ \bibnamefont {Littlewood}},\ }\href@noop {}
  {}\bibinfo {note} {(2012)} \Eprint
  {http://arxiv.org/abs/arXiv:1202.1423} {arXiv:1202.1423}
  \BibitemShut {NoStop}%
  \bibitem [{\citenamefont {O'Regan}(2012)}]{thesis}%
  \BibitemOpen
  \bibfield  {author} {\bibinfo {author} {\bibfnamefont {D.~D.}\ \bibnamefont
  {O'Regan}},\ }\Doi {10.1007/978-3-642-23238-1} {\emph {\bibinfo {title}
  {Optimised Projections for the Ab Initio Simulation of Large and Strongly
  Correlated Systems}}},\ \bibinfo {edition} {1st}\ ed.,\ \bibinfo {series}
  {Springer Theses}, Vol.\ \bibinfo {volume} {XVI}\ (\bibinfo  {publisher}
  {Springer},\ \bibinfo {address} {Berlin, Heidelberg},\ \bibinfo {year}
  {2012})\ p.\ \bibinfo {pages} {225}\BibitemShut {NoStop}%
\bibitem [{\citenamefont {Haynes}\ \emph {et~al.}(2008)\citenamefont {Haynes},
  \citenamefont {Skylaris}, \citenamefont {Mostofi},\ and\ \citenamefont
  {Payne}}]{0953-8984-20-29-294207}%
  \BibitemOpen
  \bibfield  {author} {\bibinfo {author} {\bibfnamefont {P.~D.}\ \bibnamefont
  {Haynes}}, \bibinfo {author} {\bibfnamefont {C.-K.}\ \bibnamefont
  {Skylaris}}, \bibinfo {author} {\bibfnamefont {A.~A.}\ \bibnamefont
  {Mostofi}}, \ and\ \bibinfo {author} {\bibfnamefont {M.~C.}\ \bibnamefont
  {Payne}},\ }\href {http://stacks.iop.org/0953-8984/20/i=29/a=294207}
  {\bibfield  {journal} {\bibinfo  {journal} {J. Phys.: Condens. Matter},\
  }\textbf {\bibinfo {volume} {20}},\ \bibinfo {pages} {294207} (\bibinfo
  {year} {2008})}\BibitemShut {NoStop}%
\end{thebibliography}
\end{document}